\documentstyle[12pt,aps,epsfig]{revtex}
\tightenlines
\begin{document}

\rightline{RCNP-Th/00032}

\begin{center}
{\Large \bf
Abelian Monopole and Center Vortex\\
Views at the Multi-Instanton Gas 
}

\vspace*{0.5cm}
{\bf
M. Fukushima \footnote[2]{E-mail address:~masa@rcnp.osaka-u.ac.jp} 
             \footnote[3]{present address:~Japan Atomic Energy Research Institute, 
                                                        Ibaraki 319-1195, Japan},
E.--M. Ilgenfritz \footnote[4]{E-mail address:~ilgen@rcnp.osaka-u.ac.jp}
and H. Toki \footnote[5]{E-mail address:~toki@rcnp.osaka-u.ac.jp}
}

\vspace*{0.5cm}
{\it
Research Center for Nuclear Physics (RCNP), Osaka University,\\
Mihogaoka 10-1, Ibaraki, Osaka 567-0047, Japan
}

(\today)

\begin{abstract}\baselineskip = 0.6cm
We consider full non-Abelian, Abelian and center projected lattice field 
configurations built up from random instanton gas configurations in 
the continuum. We study the instanton contribution to the $\bar{Q}Q$
force with respect to 
({\it i}) instanton density dependence, 
({\it ii}) Casimir scaling
and 
({\it iii}) whether various versions of Abelian dominance hold.
We check that the dilute gas formulation for the interaction potential 
gives an reliable approximation only for densities small compared to the
phenomenological value.  We find that Casimir scaling does not hold, 
confirming earlier statements in the literature. 
We show that the lattice used to discretize the instanton gas configurations
has to be sufficiently coarse ($a \approx 2\bar{\rho}$ compared with the 
instanton size $\bar{\rho}$) 
such that maximal Abelian gauge projection and center projection
as well as the monopole gas contribution to the $\bar{Q}Q$ force 
reproduce the non-Abelian instanton-mediated force in the intermediate 
range of linear quasi-confinement. 
We demonstrate that monopole clustering also depends critically on the 
discretization scale confirming earlier findings based on monopole blocking.
\end{abstract}
\end{center}

\noindent\hspace{1cm}{Key Word:
Instantons, monopoles, vortices, confinement
}

\noindent\hspace{1cm}{PACS number(s): 
12.38.Aw,11.15.Ha
}

\section{Introduction}
\label{sec:sec1}

A few years ago, a RCNP group including two of the present 
authors~\cite{Fukushima1,Fukushima2,Fukushima3}, has started to 
systematically study the properties of instanton systems with
respect to confinement by using multi-instanton simulations.
The instanton is a quasi-pointlike topological object in Euclidean 
space-time and, being selfdual, satisfies the Yang-Mills field 
equations~\cite{Belavin}. Mixed instanton-antiinstanton 
configurations~\cite{CDG} (gas or liquid) only locally are approximate 
solutions, but they offer the attractive possibility to solve, 
in a semiclassical-like fashion, the $U_A(1)$ problem (explain 
the large $\eta^{\prime}$ mass)~\cite{tHooft1} and to 
explain chiral symmetry breaking~\cite{Diakonov2,Shuryak1}.
(See Ref.~\cite{Schaefer} for a recent review.)
One main focus of our previous work on instantons, neglected by others, was 
to explore under which circumstances a dense, however random instanton system 
could provide confinement. In the beginning~\cite{Fukushima1} it has been pointed out that the
monopoles induced in the instanton system and detected by Abelian 
projection have the clustering property held necessary for confinement.
As for the $\bar{Q}Q$ force itself, in Ref.~\cite{Fukushima2,Fukushima3}
quantitatively the conclusion has been reached, after some refinements, 
that with the widely accepted density 
of $1~{\rm fm}^{-4}$ of instantons plus antiinstantons and an average 
instanton size of $\bar{\rho}= 0.4~{\rm fm}$, only 40 \% of the static 
$\bar{Q}Q$ force could be reproduced at distances around 
$R \approx 1~{\rm fm}$. 

This result was achieved using an instanton size distribution based 
on the idea of freezing of the strong coupling constant at large 
distance.  This assumption leads to a dimensionally dictated behavior 
of the size distribution like $dn(\rho)/d\rho \propto \rho^{-5}$ 
in the infrared. 
One should remember that the main parameter of any such model, 
the instanton density and the instanton size distribution are 
phenomenological input, beyond justification coming from a truly 
semiclassical approximation. The latter is available only for a single 
instanton~\cite{tHooft2}, however afflicted with the famous 
infrared divergence.

Phenomenologically, in the instanton liquid model instantons 
are occupying Euclidean space with a density of $n=N/V=1~{\mathrm fm}^{-4}$, 
and with an average radius $\bar{\rho}=1/3~{\mathrm fm}$ 
one gets a packing fraction 
$f=n~\bar{\rho}^4 \approx (1/3)^4$. 
For our purposes we fix only an average instanton radius $\bar{\rho}=0.4~{\rm fm}$.
We study the influence of the instanton density, considering $1~{\rm fm}^{-4}$
as the ''physical'' value. This leaves some room for the choice of the
size distribution.
 
Lattice search for instantons is {\it the method of choice}
to obtain these quantities and the size distribution relying on first 
principles. As for pure gluodynamics, {\it only} lattice studies are 
at hand presently to quantify the instanton structure of the vacuum. Over 
the last years the results have not much converged (compare the recent 
conference reports Ref.~\cite{negele,teper,garcia}). 
Different groups roughly agree on the size of instantons within a 
factor of two, e.g. $\bar{\rho}=0.3 ... 0.6~{\rm fm}$ for $SU(3)$. 
There is no agreement at all concerning the density $N/V$ which is 
strongly dependent on the procedure to remove size $O(a)$ vacuum fluctuations 
(cooling, smoothing etc.) and technical parameters (cooling steps, 
cooling radius etc.)\footnote{Sometimes the density $N/V=1~{\rm fm}^{-4}$ 
is used as a criterium to stop cooling, in this way specifying the 
remaining instanton characteristics.}. As a tendency, lattice studies 
give higher density and larger instantons than phenomenologically assumed.  
It is fair to say that only the topological susceptibility,
$\chi_{\rm top}^{\frac14}/\sqrt{\sigma}=0.45(3)$ for $SU(3)$ and
$0.50(2)$ for $SU(2)$, and the average instanton size are supported
by lattice studies. The instanton size distribution is 
much more delicate to assess. We will comment on that later.

This paper is dedicated to a critical reappraisal of the instanton
model to describe, besides other features of the Yang-Mills vacuum,
its confinement property, and we shall clearly point out the deficits
of this model. Although seemingly insufficient in quantitative respect, 
it might be interesting to see to what extent the contribution of 
instantons to the confining force depends on the $\rho$ distribution. 
Also the question whether the instanton generated $\bar{Q}Q$ force 
can be reproduced in accordance to Abelian, monopole or vortex dominance 
needs some clarification. One has to answer the question at {\it which 
scale} the model can be replaced by an effectively Abelian model with 
condensing Abelian defects. Monopole and $Z(N)$ vortex mechanisms are 
presently the leading candidates for an effective infrared description 
comprising confinement. We shall see that these descriptions are applicable 
to the instanton mediated force as well. This is expected, corroborating 
earlier work~\cite{RCNP-monop-inst} and the widely-studied interrelation 
between instantons and monopoles~\cite{Vienna,Vienna+HUB} and the newer
studies concerning the interplay of instantons and 
vortices~\cite{Negele,deForcrand}, respectively. 

Said in another way, what we want to clarify is the complementarity between 
the explicit semiclassical-like description in terms of continuous
instanton fields on one hand and the monopole and vortex aspects on
the other. The latter degrees of freedom seem to become physically 
dominant in the infrared. This is the place where, for our purposes, the 
lattice discretization appears: it is an infrared matching scale 
between the instanton picture and the gauge singularities which become manifest 
in the result of gauge fixing and Abelian projection. If the discretization 
scale is chosen too small, monopole and vortex degrees of freedom can be 
identified as well. However, a complementary description of the instanton 
mediated force can be achieved only if the matching scale is somewhere
between the size of and the distance between instantons. 

For the sake of clarity we stress that this paper is {\it not a lattice} 
study. The lattice is employed here only to enable the necessary 
coarse-graining of a continuum model. In a lattice gauge theory investigation 
the r\^ole of the discretization scale $a$ would be completely different. 
There $a$ is an ultraviolet cut-off which permits, 
at the cost of a running bare 
lattice coupling $\beta \propto 1/g^2$, not to deal {\it explicitly} with 
fluctuations of smaller and smaller wave lengths. In this case, the requirement
of scale invariance in the limit 
$a \Lambda \propto \exp\left(-\frac{6~\pi^2}{11~N_c} \beta\right) 
\rightarrow 0$ is indeed
crucial and must be confirmed for any dimensionful quantity to be physical.
Instantons can be found which are stochastically generated within the sample 
of fields.
Being lumps of gauge invariant topological density they are of immediate 
physical importance. As mentioned above, their average size is relatively 
well-defined, independently of the lattice scale $a$ (for suitable methods 
of suppressing the shortest fluctuations living on the lattice). For the 
confining Abelian defects the situation is somewhat different. 
They are identified by gauge fixing (which is practically performed on the 
lattice of scale $a$), and the scale invariance of the corresponding density, 
of the distribution of length or area etc. are controversial. Moreover, 
it is generally agreed that in order make these defects condensing they 
have to be defined with some extension (blocked monopoles or thick vortices), 
a scale which becomes decoupled, in the continuum limit, from the lattice scale
$a$. As far as instantons are discussed as a possible microscopic mechanism 
to induce Abelian defects, finding the correct matching scale is tantamount 
to define this extension.

There have been also statements in the literature~\cite{Simonov} 
criticizing the instanton contribution for violating Casimir scaling of 
the $\bar{Q}Q$ force.
While this feature of the non-perturbative force at 
intermediate distances~\cite{Bali}
is largely not understood within Yang-Mills theory
and continues to pose a problem for other models of confinement, 
we want to make clear to what extent this criticism is justified
in the case of instanton based models. 
Finally, many discussions 
on the instanton generated $\bar{Q}Q$ force are based on a dilute
gas formula worked out by Callan, Dashen and Gross~\cite{CDG-pot}. Therefore 
it is of interest to demonstrate under which circumstances the result of 
multi-instanton simulations deviates from the one-instanton approximations 
made in Ref.~\cite{CDG-pot}.
In the physical range of packings the effect of different instantons on
the Wilson loop is not expected to factorize anymore.

We believe that this study contains some lessons for other attempts 
of an explicit modeling of the QCD vacuum. The observed scale sensitivity 
of the monopole or center vortex description seems to be a more 
general feature of the complementarity between semiclassical continuum 
models for non-perturbative vacuum structure and condensing defects.
It should be remembered that the concept of Abelian and monopole dominance 
had been introduced as a property of gauge fields in the infrared
\cite{tHooft3,Ezawa+Iwasaki,Yotsuyanagi,Monopole-force1,Monopole-force2,MMP+Schilling}, not necessarily on the lattice. 
Practically, however, all evidence comes from doing the gauge fixing
and Abelian projection on the lattice for gauge fields generated by lattice
simulation. In the present work the instanton model continuum configurations 
will be discretized with the purpose to perform the same steps 
following~\cite{Monopole-force1,Monopole-force2} for the monopole part 
and~\cite{Vortex-force1,DCG,Vortex-force2} for the vortex part of the 
heavy quark force.
Explicitly calculating these contributions, it turns out that the 
discretization scale $a$ is  
an influential {\it infrared} coarse-graining parameter
and must be chosen in correspondence to the typical size and density of the 
disordering continuum non-perturbative configurations forming the vacuum. 
This is what our instanton model clearly illustrates. 

In section 2 we describe how the lattice discretization
of continuum instanton-antiinstanton configurations is done. 
In section 3 we demonstrate that the one-instanton description 
of the $\bar{Q}Q$ force breaks down already in moderately dense instanton 
gases. In the same section we study to what extent the forces for quarks 
and adjoint charges deviate from Casimir scaling.
Section 4 deals with the question how coarse-graining the  
lattice has to be chosen such that Abelian projection, the 
monopole gas and the center projection are able to reproduce 
the non-Abelian instanton-mediated force. 
In section 5 we show, for the case of monopoles, that this necessary
blocking corresponds to the percolation property held necessary for 
confinement. We summarize and conclude in section 6.

\section{Continuum Multi-Instanton Configurations \\
and their Discretization}
\label{sec:sec2}

We base our studies of the $\bar{Q}Q$ force and possible 
complementary descriptions in terms of monopoles and center 
vortices on a model which comprises the Yang-Mills vacuum as 
an ensemble of random collections of instantons and antiinstantons.
The interaction is partly taken into account in the size 
distribution given below. 

In order to fix the scale of our continuum model, we choose an 
average instanton size $\overline{\rho}=0.4~{\rm fm}$. 
The average size is most solidly defined by the profile of the 
topological density after a few cooling steps. Our choice 
realistically applies to $SU(2)$ Yang-Mills theory. 
Unless stated otherwise (when we study the density 
dependence of the instanton mediated force in section 3) 
the density is chosen $N/V = (N_{I}+N_{\bar{I}})/V=1~{\rm fm}^{-4}$
with the packing fraction $f=\bar{\rho}^4\frac{N}{V}=0.0256$. 

We adopt the sum ansatz~\cite{Diakonov1}
\begin{eqnarray}  
A_{\mu}(x)=\sum_{k} A_{\mu}^{I}(x;z_{k},\rho_{k},O_{k}) 
+\sum_{\bar{k}} A_{\mu}^{\bar{I}}(x;z_{\bar{k}},
\rho_{\bar{k}},O_{\bar{k}}),
\label{eq:S_ANS}
\end{eqnarray}  
in terms of instanton and anti-instanton solutions in the 
singular gauge where an instanton is written as
\begin{equation}
A^{I}_{\mu}(x;z,\rho,O) =  
\frac{2i~O^{ab}\bar{\eta}^{b\mu\nu}(x-z)_{\nu}\rho^{2}}
{(x-z)^{2}~[ (x-z)^{2}+\rho^{2}~] } \frac{\tau^{a}}{2} . 
\end{equation}
Here $\rho$ and $z$ denote the instanton size and the space-time position 
of the instanton center, respectively. The instanton solution 
can be rotated in color space by the color orientation matrix 
$O$. The 't~Hooft symbol $\bar{\eta}^{b\mu\nu}$ is defined as 
$\bar{\eta}^{b\mu\nu} \equiv \varepsilon^{b\mu\nu} 
(1-\delta^{4\mu}\delta^{4\nu})-\delta^{b\mu}\delta^{4\nu}+
\delta^{b\nu}\delta^{4\mu}$. The anti-selfdual solution 
$A^{\bar{I}}_{\mu}$ is obtained replacing $\bar{\eta}^{b\mu\nu}$ 
by $\eta^{b\mu\nu}\equiv (-1)^{\delta^{\mu 4}+\delta^{\nu 4}} 
\bar{\eta}^{b\mu\nu}$. The instanton solutions have several 
gluonic collective modes related to variations of parameters 
like size and position (five collective degrees of freedom). 
For pure $SU(2)$ gauge theory, the color orientation matrix 
is characterized by 3 parameters (the Euler angles). For 
$N_c$ colors the number of collective parameters (and gluonic 
zero modes) generalizes to $4N_{c}$. 

Actually, we generate the ensemble of instantons and 
anti-instantons by randomly placing $z_{k}$ in a 4-dimensional 
Euclidean continuum box. The (adjoint) color orientations $O_{k}$ 
are taken randomly with the Haar measure, and the instanton sizes 
$\rho_{k}$ are sampled according to the following size distribution : 
\begin{equation}
\frac{dn(\rho)}{d\rho} = \alpha~\rho^{b-5} \exp(-\beta\rho^2/
\overline{\rho^2})
\label{eq:SizeDist}
\end{equation}
with $b=11N_{c}/3$. Here $\alpha$ and $\beta$ are fixed by 
normalizing to the space-time density as $\int^{\infty}_{0} 
dn(\rho) = N/V$ and the average size $\int^{\infty}_{0}  
\rho~dn(\rho) = \bar{\rho}~N/V $. 
In the explicit configurations, the instanton number 
$N_{I}$ is taken equal to that of the antiinstantons $N_{\bar{I}}$.

This form of the size distribution was originally established in a mean-field 
treatment of color-averaged interactions~\cite{Diakonov1}.
Averaged over orientations, the interaction of equal-sign and opposite-sign
pairs is repulsive, $S_{\rm int} \propto \rho_1^2~\rho_2^2$. In the
mean-field approximation for the one-instanton distribution, this results 
in a suppression exponentially in $\rho^2$. 

As for the lattice evidence for such an instanton size distribution we 
refer to the systematic problems mentioned in the Introduction.
One should keep in mind that instanton finding algorithms depend on
cooling or smoothing. Huge instantons distinguished by a 
very weak field strength could become washed out within the first steps of 
cooling~\cite{Ilgenfritz+Thurner} together with the noise. 
Smith and Teper~\cite{UKQCD} have reported an 
instanton $\rho$ distribution from quenched $SU(3)$ lattice ensembles 
resembling an exponential cutoff at large $\rho$.
The size distribution found by Hasenfratz and Nieter~\cite{Hasenfratz+Nieter} 
has been interpreted by Shuryak~\cite{Higgs} in an exponentially suppressed
parameterization in terms of a dual Higgs mechanism.
There are also other warnings, motivated by the wealth of semiclassical
configurations known today~\cite{vBaal+Margarita}, not too early to take 
for granted the size cut-off as seen in the lattice studies 
of Refs.~\cite{UKQCD,Hasenfratz+Nieter}.

Very recently, in Ref.~\cite{Muenster}, a distribution like (\ref{eq:SizeDist})
has been demonstrated to emerge, in dimension $d>2$, from a very general model of
soft (inflatable) spheres with excluded-volume interaction which inherits 
from QCD not more than the semiclassical perturbative instanton fugacity 
proportional to $\rho^{b-5}$. In the grand canonical Monte Carlo simulation 
of this system the $\rho$ dependent fugacity has been used in conjunction 
with the excluded-volume constraint.

Because the instantons are randomly placed inside the box, it might be 
interesting to know how frequently instantons are strongly overlapping, 
which we define by $\rho_1~\rho_2/\Delta_{12}^2 > 1$ (where $\Delta_{12}$ 
is the Euclidean distance). For the size distribution (\ref{eq:SizeDist})
and the physical density of $1~{\mathrm fm}^{-4}$,
the number $N_{<}$ of strongly overlapping pairs (including all possible 
pairings between instantons and antiinstantons) relative to the total 
number $N=N_{+}+N_{-}$ of instantons and antiinstantons amounts to 
$N_{<}/N=0.072$. This fraction actually depends on the width of the size 
distribution. For instance, for a sharp distribution
$\frac{dn(\rho)}{d\rho} \propto \delta(\rho-\bar{\rho})$ 
one would find, with the same density and $\bar{\rho}$, 
a smaller fraction $N_{<}/N=0.056$ of closely packed pairs.
We are aware of the problem that the linear superposition ansatz
should be replaced by some better ansatz (ratio ansatz)  
for close instanton-instanton pairs and by the streamline 
ansatz for instanton-antiinstanton pairs at high densities. 
For the observed percentage of close pairs in the physical range of
densities, we have neglected this potential complication. This improvement 
would become more important with an algebraically falling size distribution. 

In our actual calculation, we cover the random multi-instanton 
configuration by a lattice similar to Ref.~\cite{Fukushima3}. 
For the calculation of the fully non-Abelian force the discretization 
actually would not be necessary. In fact, we choose a sufficiently fine 
lattice spacing of $a=0.05~{\rm fm}$ (this should be compared with the 
average instanton size of $\bar{\rho}=0.4~{\rm fm}$). For this 
case (of the finest lattice) the link variables are given simply 
as $U_{\mu}(x) = \exp [ia A_{\mu}(x+\frac12\hat{\mu}a)] $ using 
the gluon field of the multi-instanton system on the mid-point 
of the link $l=\{x,\mu\}$. In connection with Abelian gauges and 
Abelian projections, we will apply a coarser discretization 
scale. Then the $SU(2)$ link variables are constructed by integrating 
the vector potential defined on the continuum space as
\begin{eqnarray}
U_l&=&U_{\mu}(x)~=~
{\rm P}~\exp \Bigm[ \int^{x+\hat{\mu}a}_{x} dx^{\prime}_{\mu} 
A_{\mu}(x^{\prime}) \Bigm] \nonumber\\
&=&{\rm P}~\prod^{l}_{j = 1}  
\exp \Bigm[ \tilde{a}~A_{\mu}(x+(j\!-\!\frac12)~\hat{\mu}\tilde{a} 
) \Bigm].
\end{eqnarray}
Here, the path from $x$ to $x+\hat{\mu}a$ has been subdivided 
into smaller segments with step size of $\tilde{a}=a/l$, and 
the above path ordered exponential has been calculated as a 
product over finer links defined on these segments. Actually, 
we take $\tilde{a} = 0.05~{\rm fm}$ as the segment size.
This construction becomes increasingly important when the 
discretization scale of the lattice gets comparable with or
larger than the average instanton size. 

Periodicity of the lattice gauge field configurations has been 
enforced by placing the 4-dimensional Euclidean box of size 
$V=(4.8~{\rm fm})^{4}$ (to be covered by the lattice) into 
a bigger box. The 3-dimensional boundary in each of the eight 
directions is extended to a 4-dimensional slab of thickness $1.2~{\rm fm}$ 
continuing the basic box. In each of these slabs copies (phantom
instantons) are placed of instantons which are near to the opposite 
boundary. 
These are also included into the sum (\ref{eq:S_ANS}) representing
the continuum vector potential. Then, along the 
links restricted to the basic box the above construction is performed. 

\section{The $\bar{Q}Q$ Potential from the Random Instanton Liquid \\
and the Dilute Gas Limit} 
\label{sec:sec3}

The effect of instantons on the heavy quark potential has been 
first discussed in Ref.~\cite{CDG-pot} by using the dilute 
gas approximation. There a formula 
has been derived according to which the static potential could 
be written as \begin{equation}
V(R) = 2 \int d\rho \frac{dn(\rho)}{d\rho}~\rho^3~W\left(\frac{R}{\rho}\right), 
\label{eq:CDG}
\end{equation}
where $dn(\rho)/d\rho$ is the respective instanton size 
distribution and the factor two refers to instantons plus 
antiinstantons. Here, a dimensionless potential $W(R/\rho)$ ($R=|\vec{r}|$) 
is defined by integration over the instanton position $\vec{x}$ 
\begin{equation}
W\left(\frac{R}{\rho}\right) = -\frac{1}{2 \rho^{3}} \int  d^{3}x~{\mathrm tr}
~\bigm[ U(\vec{r}-\vec{x})U^{\dagger}(-\vec{x})-1\bigm].
\label{eq:auxiliary_function}
\end{equation}
The fundamental static charges ($j=\frac12$) are represented 
by static trajectories positioned at $\vec{0}$ and $\vec{r}$ in 3-space 
(with $R=|~\vec{r}~|$). The color matrix $U(\vec{r}-\vec{x})$ 
represents the path ordered exponential along one of the 
infinite trajectories traversing the instanton and is given 
as $U(\vec{r}-\vec{x}) = \exp [ i~\pi~\vec{\tau} \cdot
(\vec{r}-\vec{x})/ \bigm[ (\vec{r}-\vec{x})^{2}
+\rho^{2}\bigm]^{1/2}]$. It depends only on the instanton 
size $\rho$ and the shortest distance from the trajectory 
to the instanton center. Notice that the instanton orientation 
does not matter. That part of integration is normalized to unity.
Replacing the trace $\frac12 {\mathrm tr}(U_1~U_2^{\dagger})$ in 
(\ref{eq:auxiliary_function}) by the trace in the adjoint representation, 
$\frac13 {\mathrm tr}_{{\mathrm ad}} (U_1~U_2^{\dagger}) 
= \frac13 \left( ({\mathrm tr} (U_1~U_2^{\dagger}) )^2 - 1 \right)$ 
one can extend (\ref{eq:CDG}) to the adjoint ($j=1$) charges.                                                                                                                                                                                                             
For the comparison in this section we adopt the instanton 
size distribution in the form of eq.~(\ref{eq:SizeDist}) and 
confront the dilute gas result (\ref{eq:CDG}) with  
multi-instanton simulation results where the same distribution 
has been used for sampling of the $\rho_k$'s. In this section, the total 
density $N/V$ is a free parameter.

\begin{figure}[h]
\begin{center}
\begin{minipage}[hbt]{6cm}
\includegraphics[width=6cm]{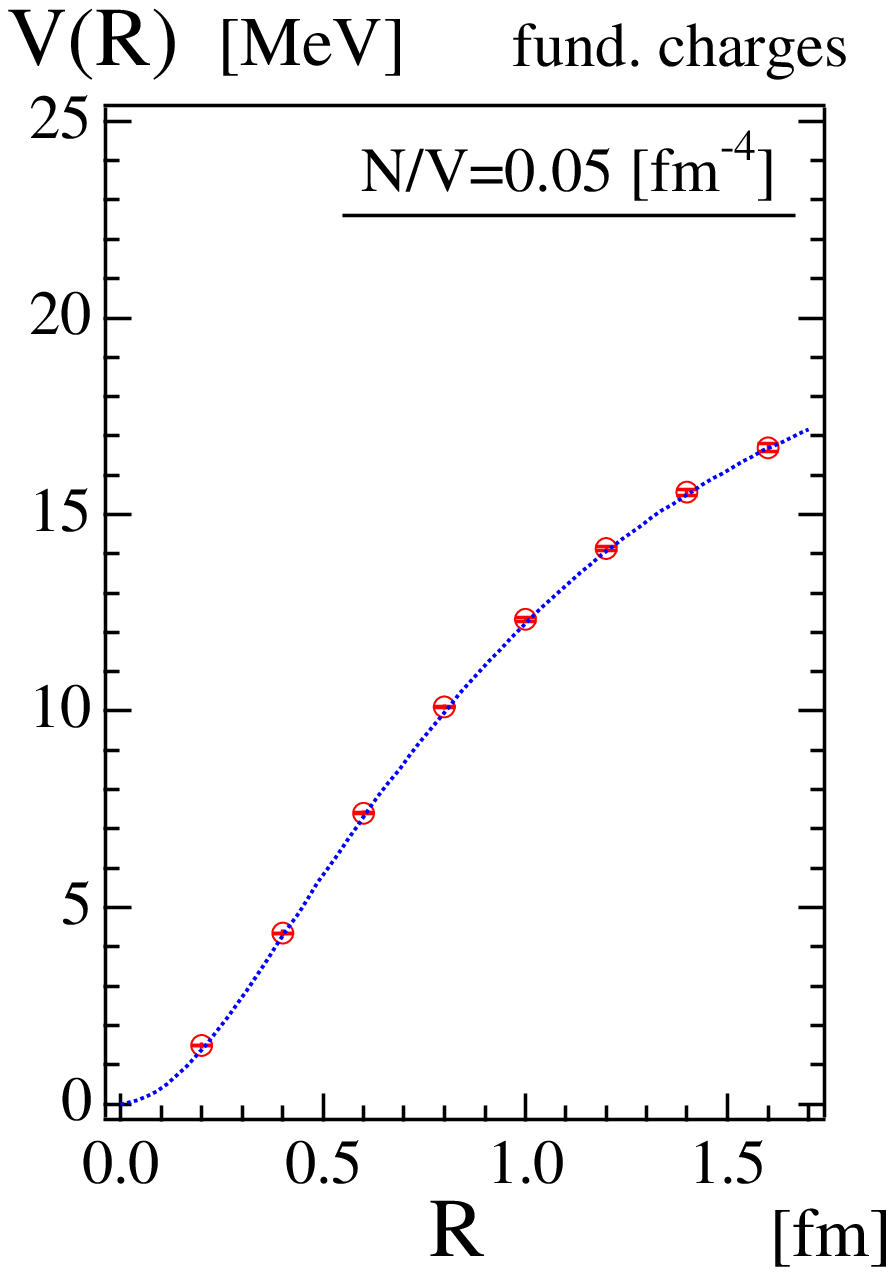}
\vspace*{-1.0cm}
\center{(a-i)}
\end{minipage}
\begin{minipage}[hbt]{6cm}
\includegraphics[width=6cm]{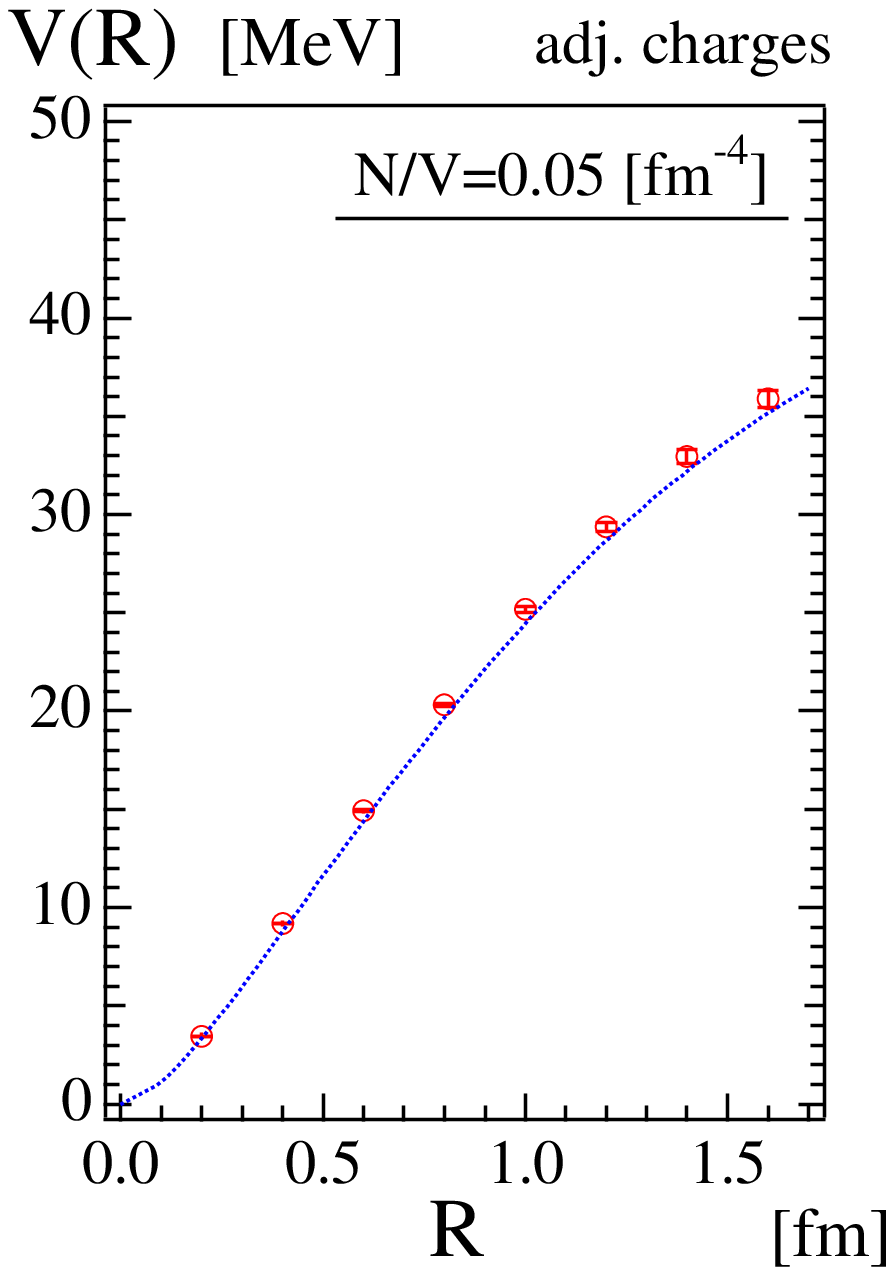}
\vspace*{-1.0cm}
\center{(a-ii)}
\end{minipage}
\end{center}

\vspace*{0.8cm}

\begin{center}
\begin{minipage}[hbt]{6cm}
\includegraphics[width=6cm]{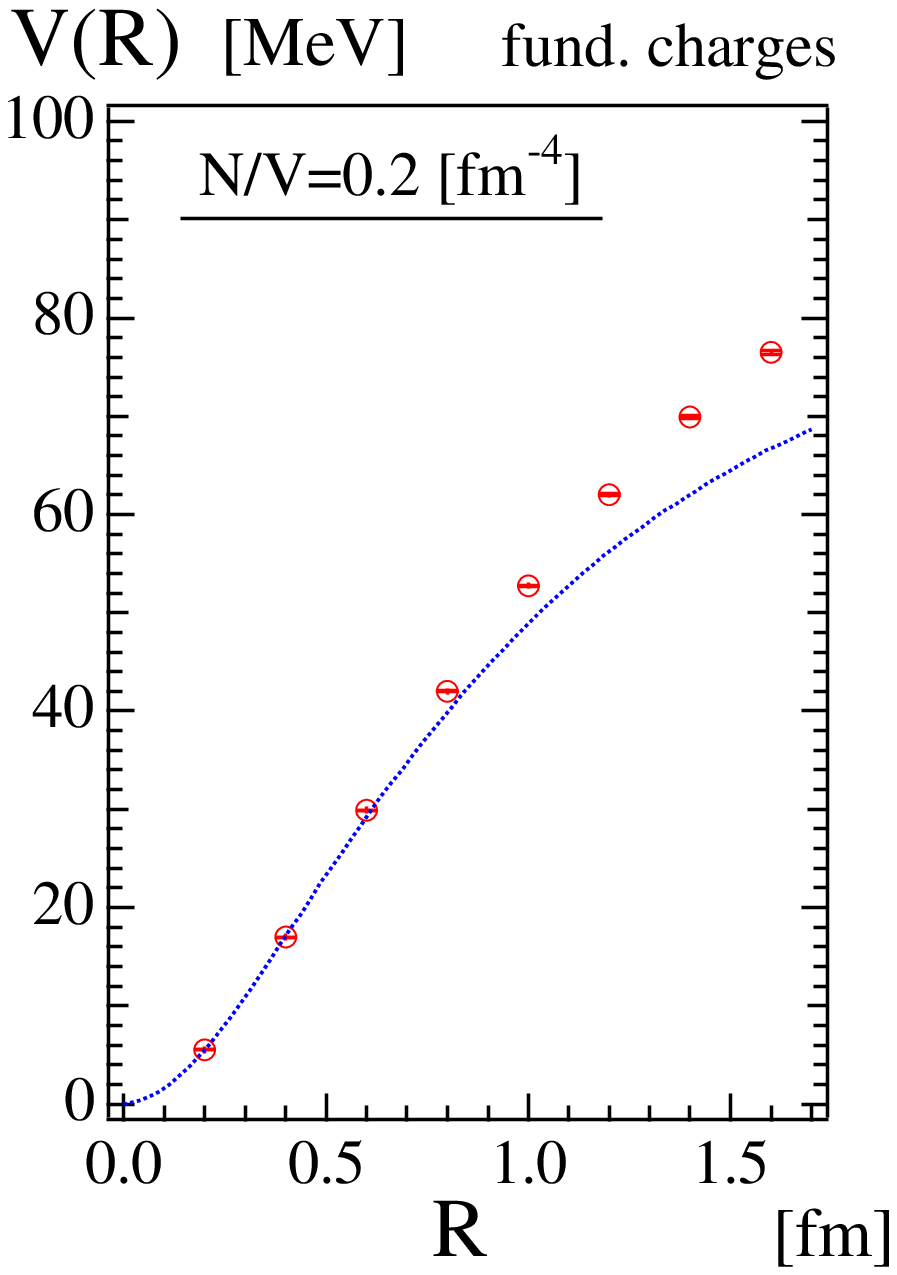}
\vspace*{-1.0cm}
\center{(b-i)}
\end{minipage}
\begin{minipage}[hbt]{6cm}
\includegraphics[width=6cm]{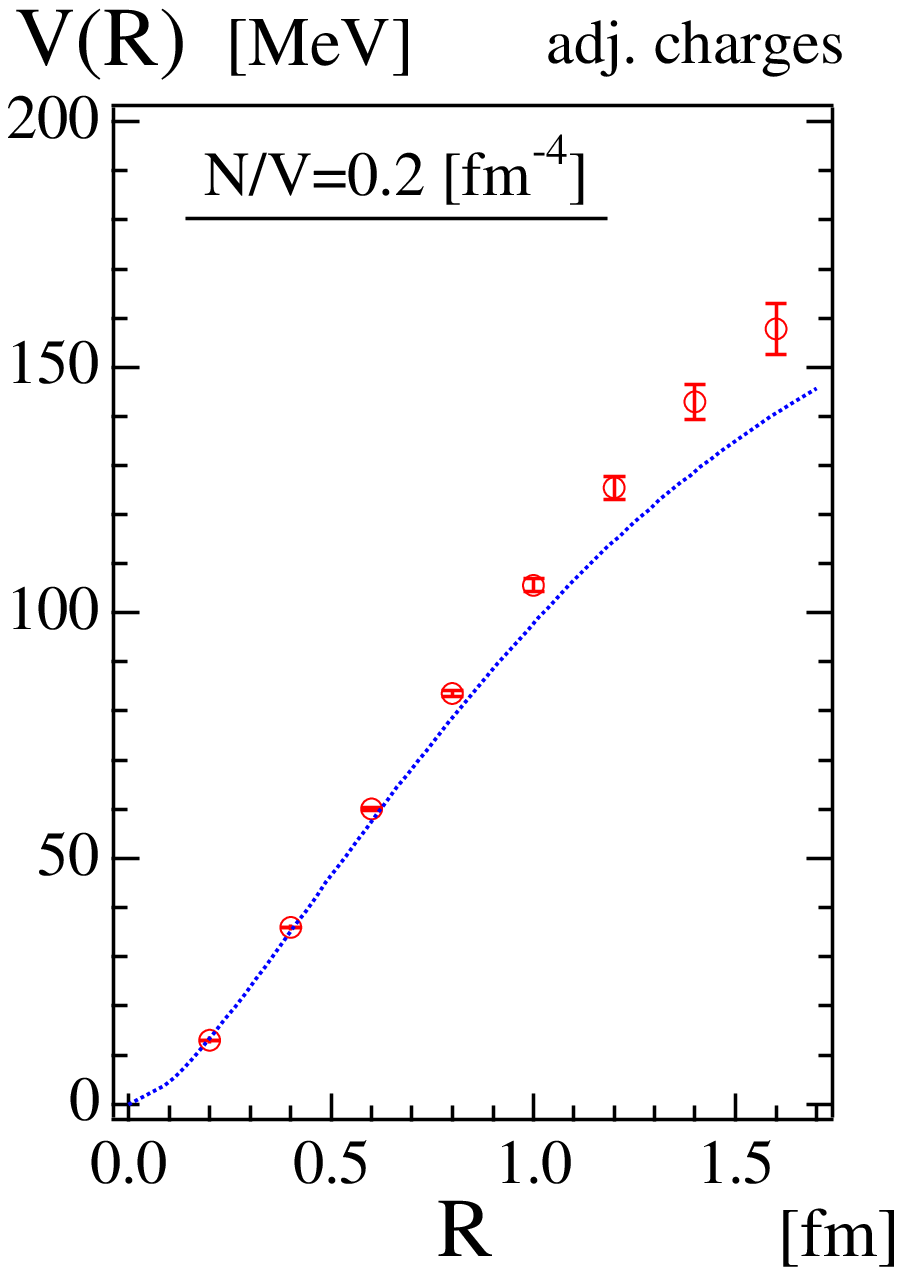}
\vspace*{-1.0cm}
\center{(b-ii)}
\end{minipage}
\end{center}

\vspace*{0.8cm}

\begin{center}
\begin{minipage}[hbt]{6cm}
\includegraphics[width=6cm]{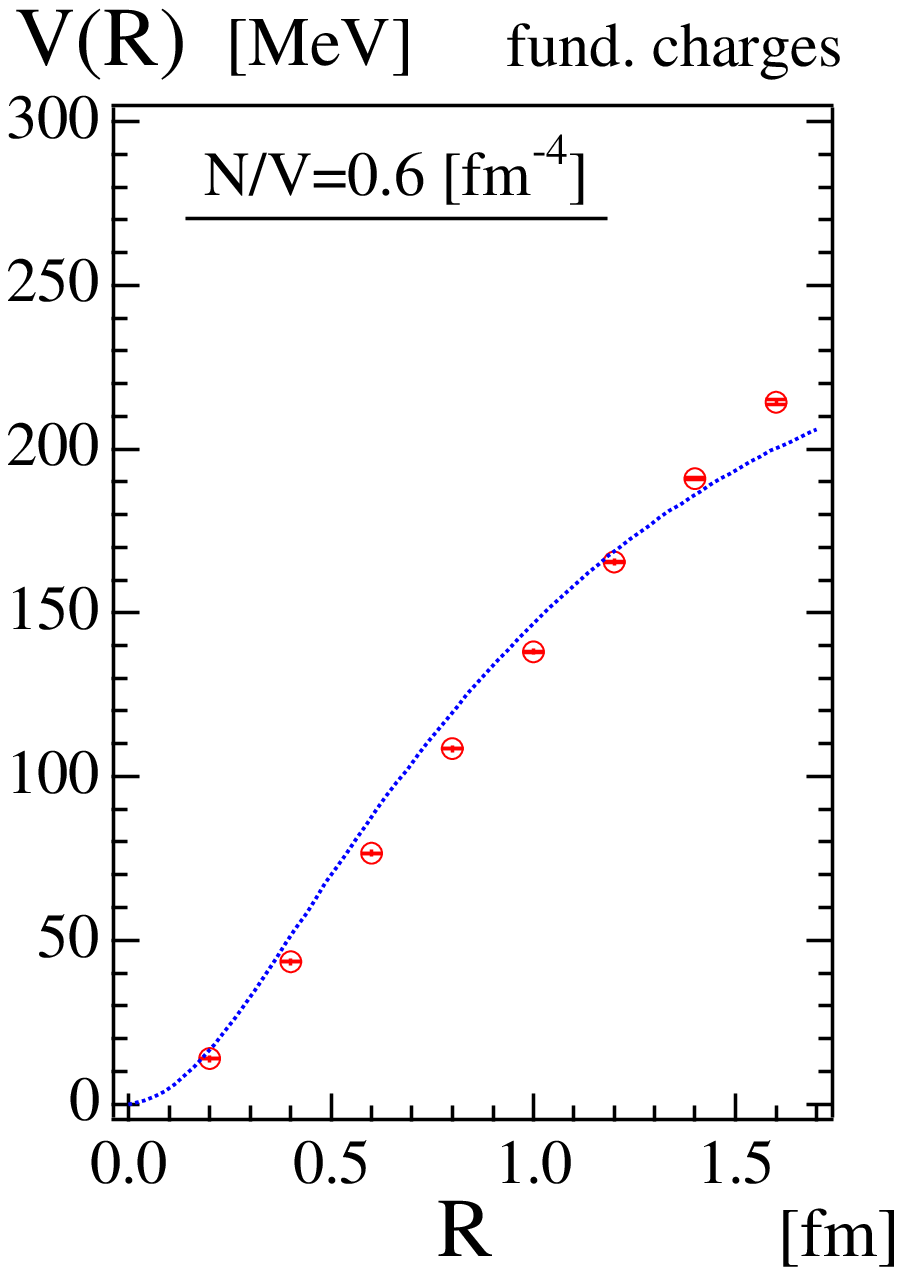}
\vspace*{-1.0cm}
\center{(c-i)}
\end{minipage}
\begin{minipage}[hbt]{6cm}
\includegraphics[width=6cm]{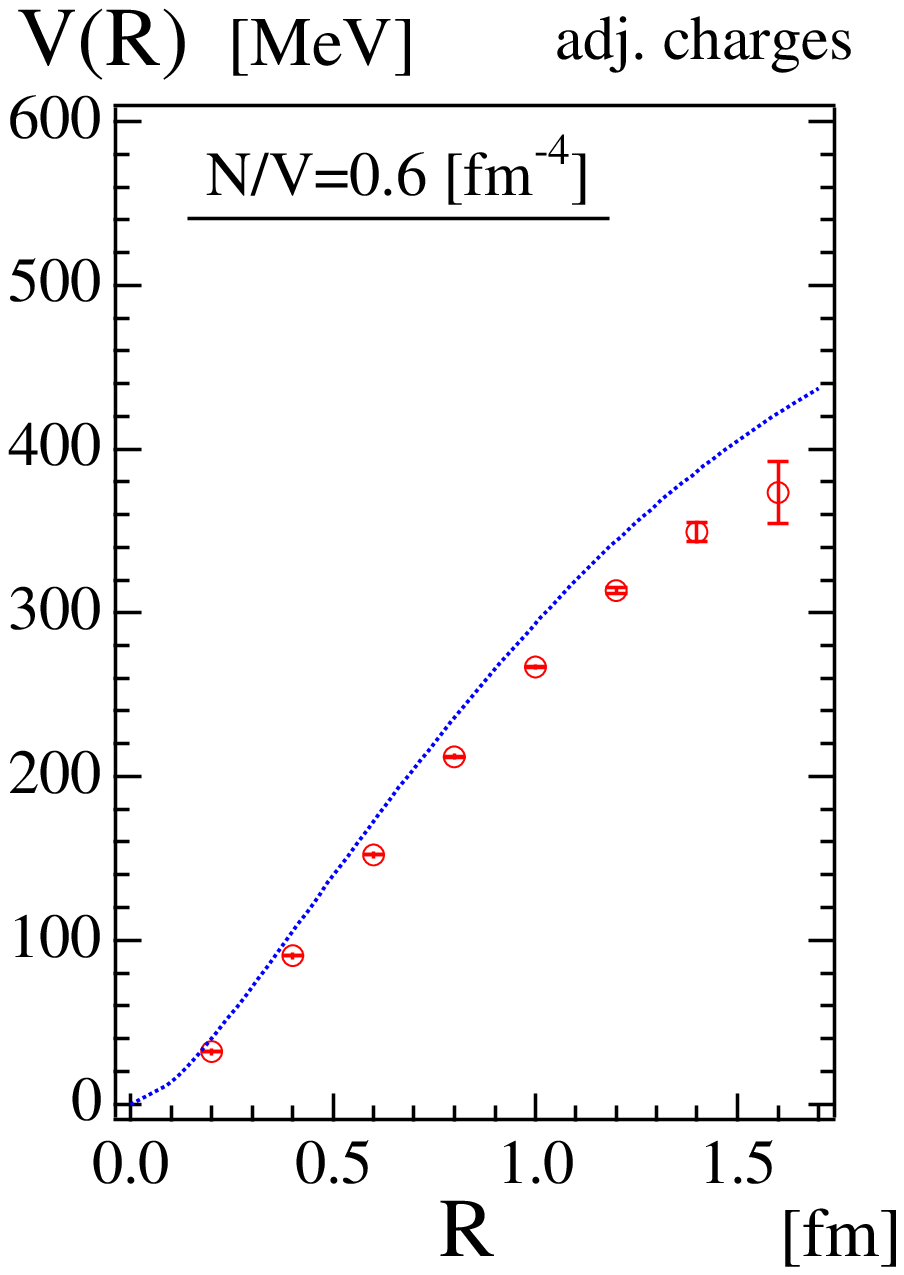}
\vspace*{-1.0cm}
\center{(c-ii)}
\end{minipage}
\end{center}

\vspace*{0.8cm}

\begin{center}
\begin{minipage}[hbt]{6cm}
\includegraphics[width=6cm]{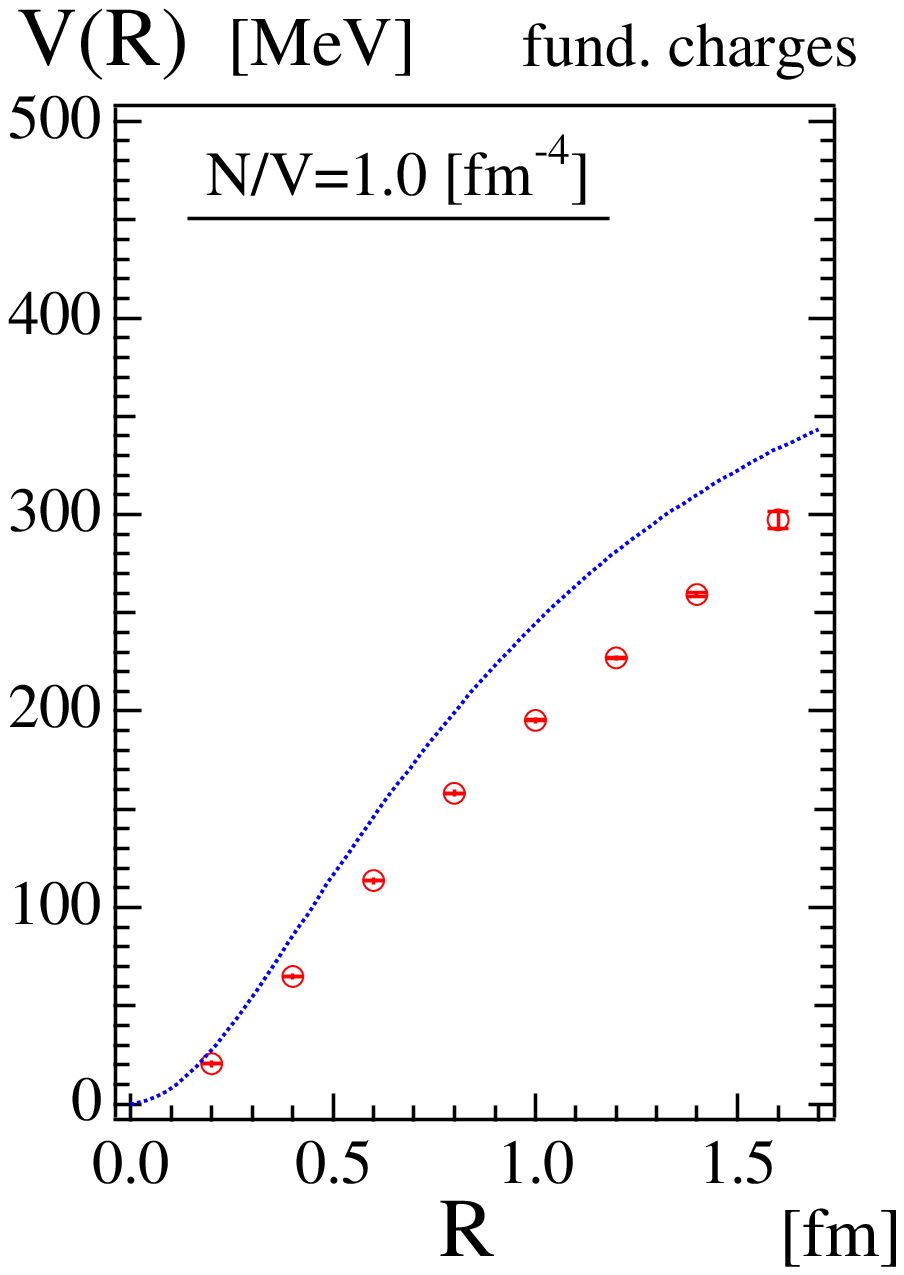}
\vspace*{-1.0cm}
\center{(d-i)}
\end{minipage}
\begin{minipage}[hbt]{6cm}
\includegraphics[width=6cm]{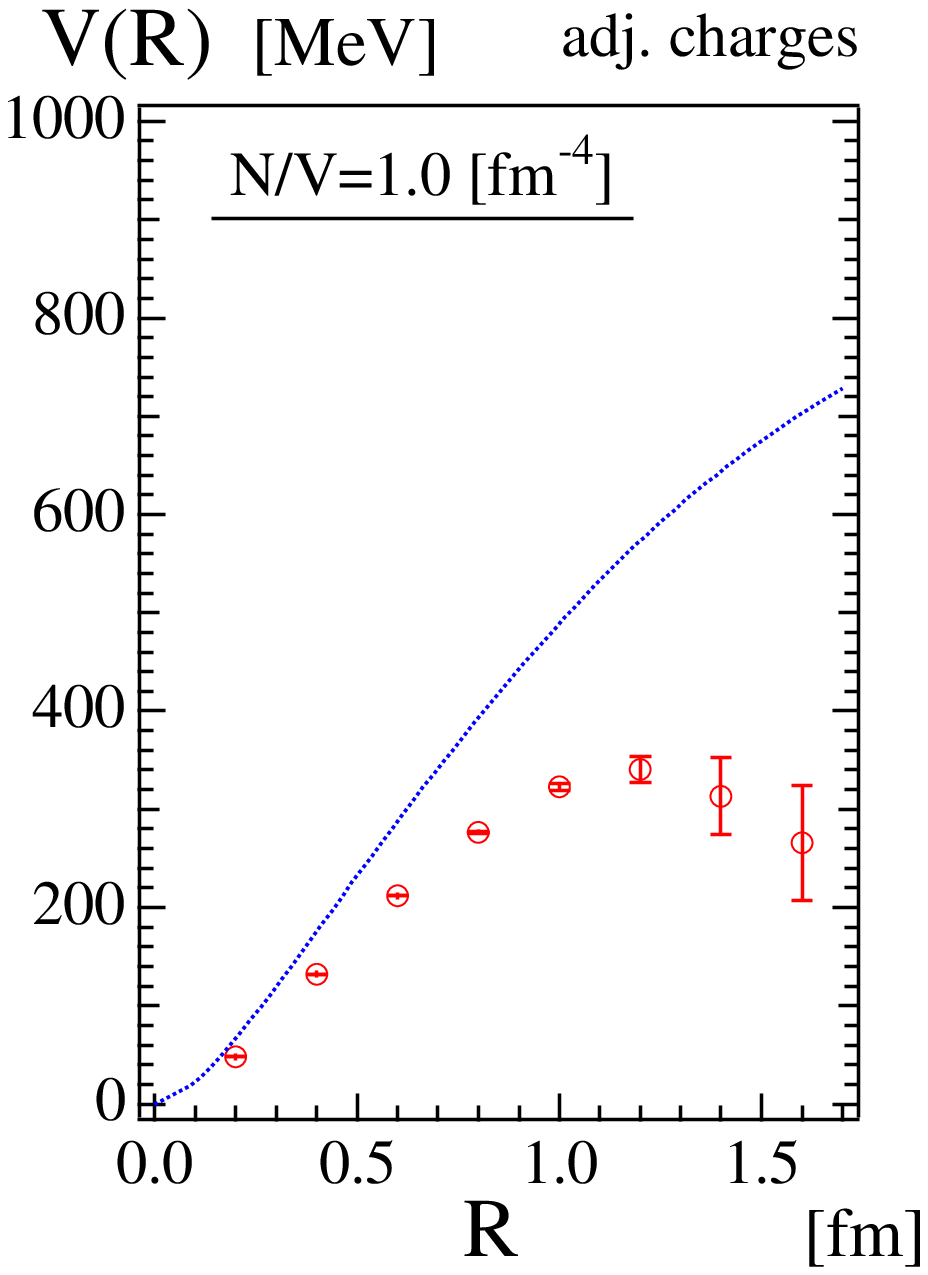}
\vspace*{-1.0cm}
\center{(d-ii)}
\end{minipage}
\end{center}

\caption{
The static non-Abelian potential derived 
from simulations of the random instanton 
liquid and from a dilute gas (one-instanton) 
approximation 
for fundamental (i) and adjoint (ii) charges
The comparison is made for instanton densities 
$N/V=0.05,~0.2,~0.6$ and $1.0~{\rm fm}^{-4}$.
The curves describe the dilute gas approximation,
the data points simulation results.}
\label{fig:PCDG}
\end{figure}

The potential (for fundamental and adjoint charges) obtained from 
eq.~(\ref{eq:CDG}) is simply proportional to the density $N/V$ as shown in 
Fig. \ref{fig:PCDG} by the solid lines. 
The data symbols in Fig. \ref{fig:PCDG} denote the results 
of an uncorrelated multi-instanton simulations. 
The comparison is made for the instanton densities 
(a) $N/V=0.05~{\rm fm}^{-4}$,  
(b) $N/V=0.20~{\rm fm}^{-4}$,  
(c) $N/V=0.60~{\rm fm}^{-4}$, and 
(d) $N/V=1.00~{\rm fm}^{-4}$.
The results of the simulation are based on a statistics of 
(a) 100,
(b) 300,
(c) 500, and
(d) 1000 configurations respectively. 
The left panels show the potential between fundamental charges,
while the right panels between adjoint charges. 

The simulation data have been obtained, using the lattice 
discretization of the random instanton-antiinstanton 
configurations as described above, from expectation values 
of Wilson loops 
\begin{equation}
\langle W_{full}(R,T) \rangle 
= \langle {\rm Tr} \prod_{l \in C} U_{l} \rangle 
\end{equation}
where the contour $C$ is a rectangular closed path of size 
$R \times T$ on the finest lattice of $a=0.05~{\rm fm}$, 
a spacing almost 10 times smaller than the average instanton 
size. The non-Abelian potential
\begin{equation} 
V(R) = - \frac{1}{a}~\log  
\frac{\langle W_{\rm full}(R,T) \rangle}
     {\langle W_{\rm full}(R,T-1)\rangle} 
\end{equation}
has been constructed from the discrete logarithmic time 
derivative.
 
We see that eq. (\ref{eq:CDG}) is only approximate and
applicable only for a system of widely separated uncorrelated 
instantons and/or antiinstantons. Then the effect of single
instantons exponentiates in a Wilson loop at finite temporal 
extension $T$. 
Only at very small distance, where overlapping of instantons 
can be neglected, simulation and analytical formula agree 
in the quadratic behavior of the potential.

Already for a moderate density of $N/V=0.2~{\mathrm fm}^{-4}$
deviations are remarkable which develop differently with higher 
density for fundamental and adjoint charges.  
From the comparison we conclude that the dilute gas formula
for fundamental and adjoint charges 
is valid only up to a density of $0.05~{\rm fm}^{-4}$ 
which corresponds to a packing fraction 
$f= 1.3~\times 10^{-3}$.
Corrections become visible first at large distances of 
$R \approx 1.5~{\rm fm}$. We show this for the density 
$0.2~{\rm fm}^{-4}$ corresponding to a packing fraction 
$f= 5.2~\times 10^{-3}$.
The resulting potential from the multi-instanton 
simulation continues to rise where the dilute-gas formula 
starts already leveling off. 

\begin{figure}[h]
\begin{center}
\includegraphics[width=12cm]{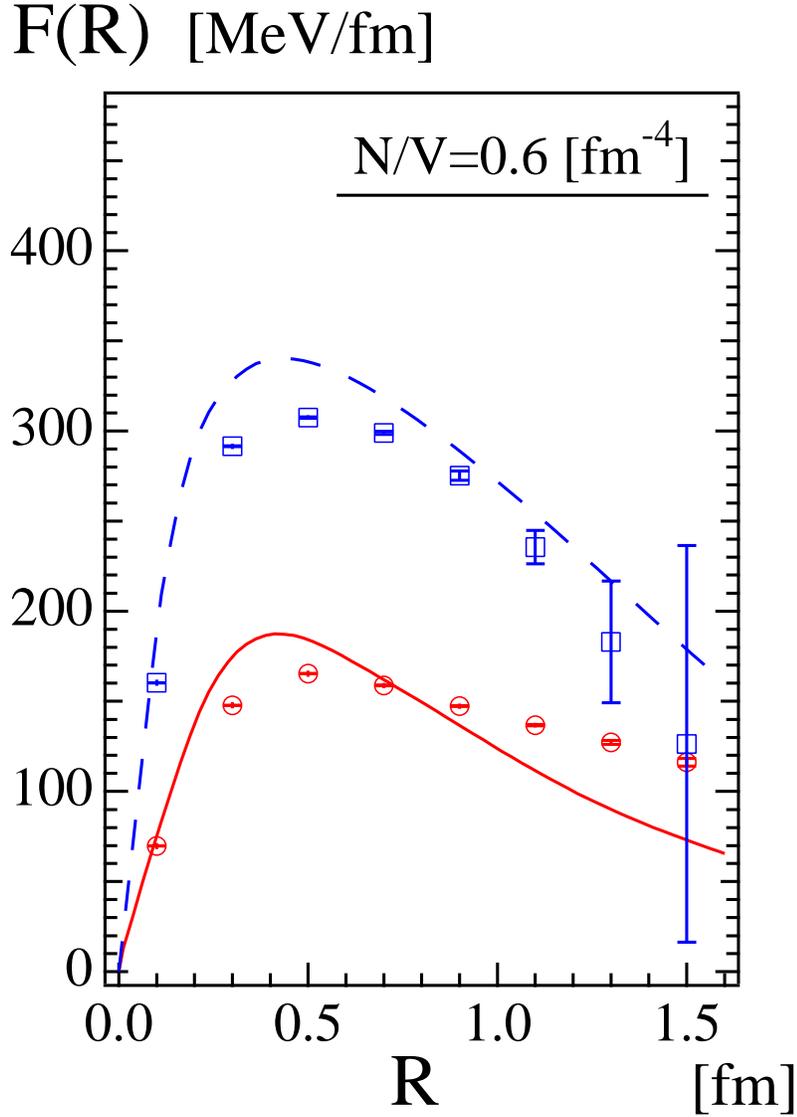}
\vspace*{-0.3cm}
\end{center}
\caption{
The static non-Abelian force derived 
from simulations of the random instanton 
liquid and from a dilute gas (one-instanton) 
approximation for fundamental (circle) and adjoint (square) 
charges. The comparison is made for an instanton density
$N/V=0.6~{\rm fm}^{-4}$.
The curves and the dotted curves describe the 
dilute gas approximation for fundamental and 
adjoint charges, respectively.}
\label{fig:FCDGnew}
\end{figure}

For the comparison of the instanton contributions to the forces 
between ''quark'' and ''antiquark'' in the fundamental and adjoint 
representation, respectively, it is more illuminating to consider 
the derivatives of the potentials shown in Fig. \ref{fig:PCDG}.
We are interested now in a instanton gas of realistic
density and show in Fig. \ref{fig:FCDGnew} the result of simulations 
for $N/V=0.6~{\rm fm}^{-4}$ only, together with the dilute gas result.
The ratio between the adjoint and the fundamental charge is
$< 2$ at all distances and is closest to the Casimir ratio, 
$\frac{9}{4}$ at $R = 0.5~{\rm fm} \approx \bar{\rho}$
where the instanton contribution itself is maximal. The ratio
becomes quite small as the distance increases. This tendency is
seen for the other densities, too.

\begin{figure}[h]
\begin{center}
\begin{minipage}[hbt]{6cm}
\includegraphics[width=6cm]{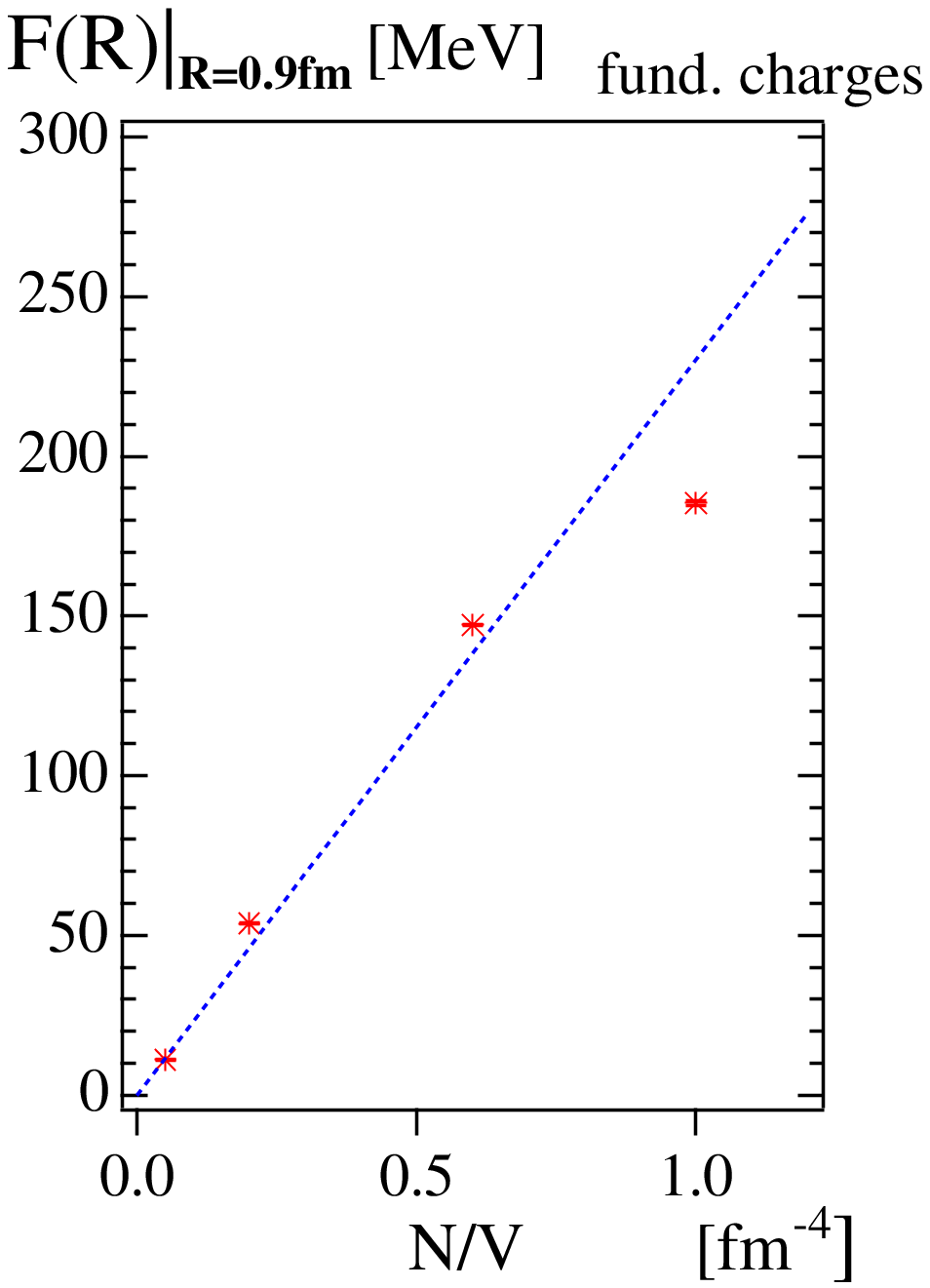}
\vspace*{-0.3cm}
\center{(a-i)}
\end{minipage}
\begin{minipage}[hbt]{6cm}
\includegraphics[width=6cm]{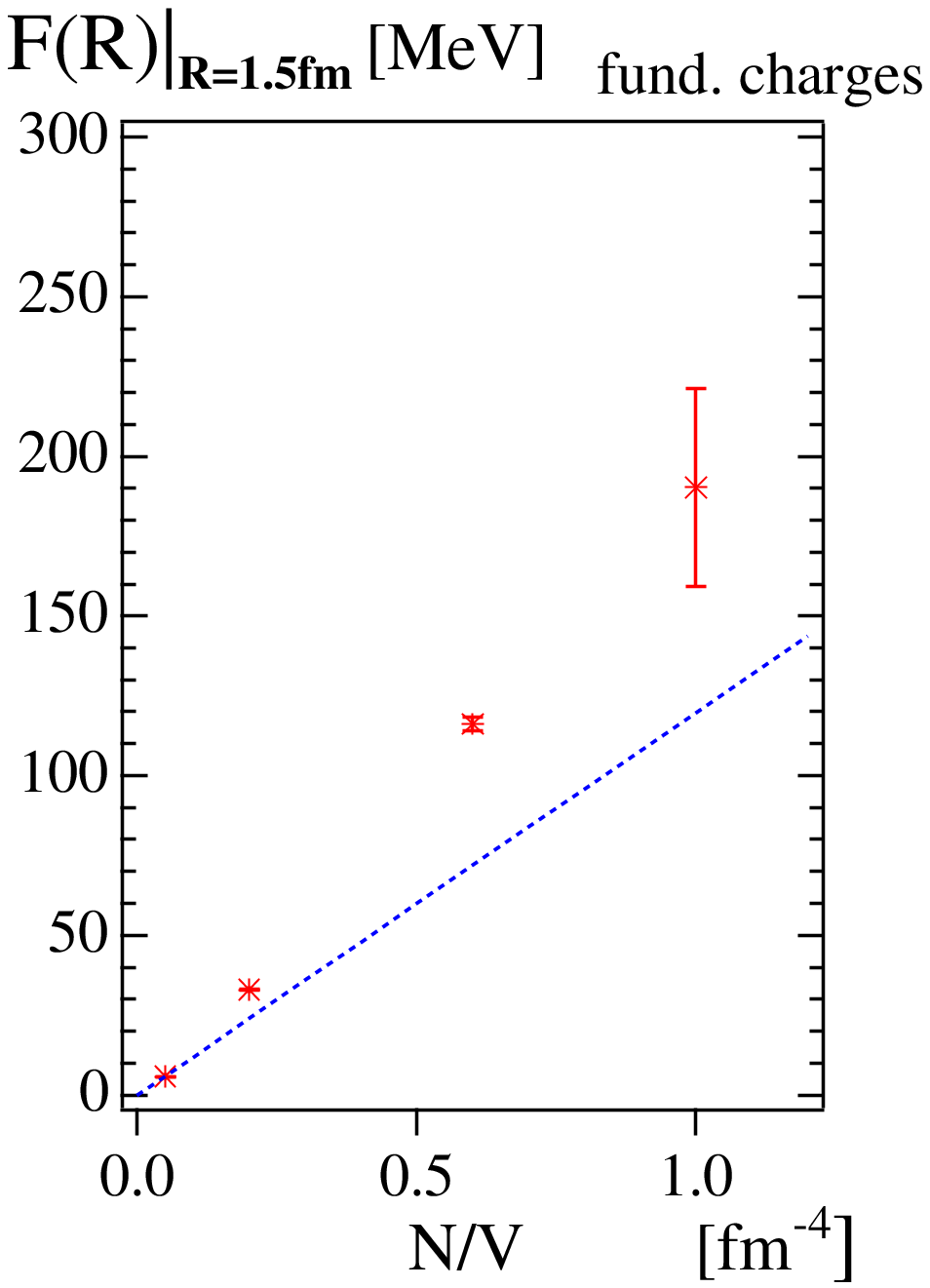}
\vspace*{-0.3cm}
\center{(a-ii)}
\end{minipage}

\begin{minipage}[hbt]{6cm}
\includegraphics[width=6cm]{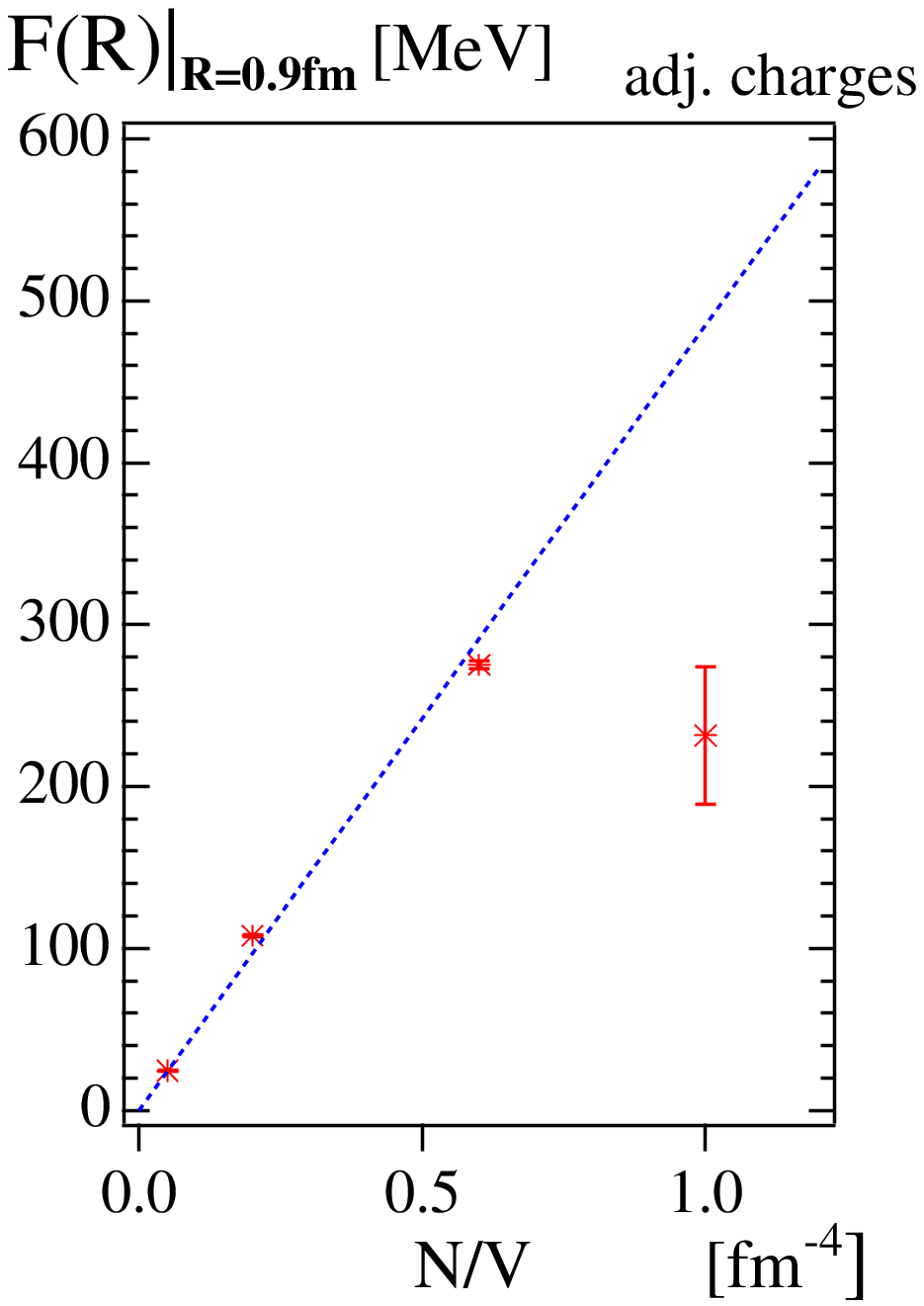}
\vspace*{-0.3cm}
\center{(b-i)}
\end{minipage}
\begin{minipage}[hbt]{6cm}
\includegraphics[width=6cm]{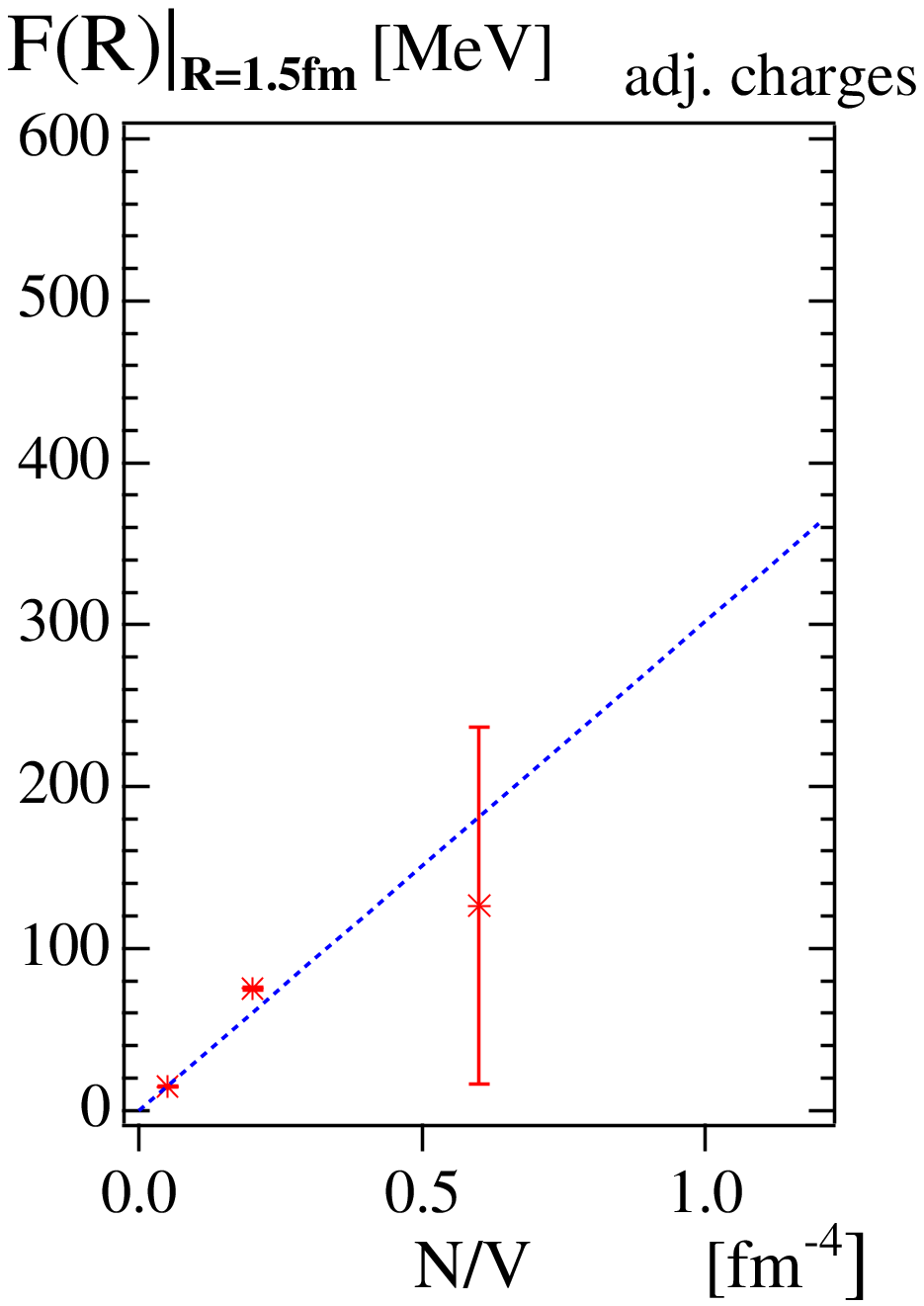}
\vspace*{-0.3cm}
\center{(b-ii)}
\end{minipage}
\end{center}
\caption{
The density dependence of the forces at
distance $R = 0.9~{\rm fm}$  and  $R = 1.5~{\rm fm}$
for (a) fundamental and (b) adjoint charges.
The solid curves describe the dilute gas approximation
and the data points correspond to the simulation results.}
\label{fig:DD}
\end{figure}

In Fig. \ref{fig:DD}
we show the density dependence of the $\bar{Q}Q$ force at two distances,
$R=0.9~{\mathrm fm} \approx 2~\bar{\rho}$ and $R=1.5~{\mathrm fm}$, 
above (a) for fundamental, below (b) for adjoint charges. It is interesting to see 
for the smaller distance that both forces rise with the instanton density $N/V$ 
almost linearly, with a slope compatible with the dilute gas (one-instanton) 
approximation, up to $N/V \sim 0.6~{\rm fm}^{-4}$. 
The forces drop down significantly at higher density, as shown by simulations 
at $N/V=1~{\rm fm}^{-4}$, in the case of fundamental charges relatively to the 
dilute gas extrapolation, while in the case of adjoint charges even absolutely.
A similar behavior is obtained for all $R < 1~{\rm fm}$. 
On the other hand, for $R > 1~{\rm fm}$ the behavior is different.
We show this for $R \sim 1.5~{\rm fm}$. The force in the fundamental
representation rises almost linearly in the density with a slope exceeding 
the dilute gas approximation by $\approx 50$ \%, while the force in the adjoint 
representation drops already at small density below the dilute gas result.

\section{Various Abelian Projections \\
of Multi-Instanton Configurations}
\label{sec:sec4}

In this section, we want to clarify how concepts like 
Abelian dominance, monopole dominance, center dominance can be applied
to semiclassical-like multi-instanton configurations. We shall compare 
the fully non-Abelian force discussed in section 2 with the results 
obtained after standard techniques of gauge fixing and Abelian projection 
have been applied to the multi-instanton fields and the respective
contributions to the static $Q\bar{Q}$ force have been evaluated.
Technically, as a coarse-graining device, the continuum configurations
are latticized, and the results will depend critically on the lattice 
spacing compared with the average instanton size. 

For this study the periodic Euclidean box of size 
$V=(4.8~{\rm fm})^{4}$ is discretized with four different lattice spacings, 
$a= 0.2~{\rm fm}$, 
$a= 0.4~{\rm fm}$,  
$a= 0.6~{\rm fm}$, and 
$a= 0.8~{\rm fm}$, corresponding to lattices 
$24^4$, 
$12^4$, 
$8^4$ and 
$6^4$, respectively. In all cases, we have chosen
$\tilde{a} = 0.05~{\rm fm}$ as the segment size (integration step)
to construct the links (path ordered exponential) of the lattice configuration.
How periodicity is enforced has been described in section 2. 

First, we consider the maximally Abelian gauge~\cite{Kronfeld} (MAG), 
which is defined by maximizing the functional
\begin{equation}
R_{MA} =  \sum_{x,\mu}~{\rm tr}~
[U_{\mu}(x)~\tau^{3}~U^{\dagger}_{\mu}(x)~\tau^{3}] 
\end{equation}
with $U_{\mu}(x)=U_{\mu}^{0}(x)+i\tau^{i}U_{\mu}^{i}(x)$.
In the MA gauge, the $SU(2)$ link variable $U_{\mu}(x)$ is 
decomposed as
\begin{eqnarray}
&&U_{\mu}(x)
= M_{\mu}(x)u_{\mu}(x) \nonumber \\
&=& 
\left(
        \begin{array}{cc}
\sqrt{1-\mid c_{\mu}(x)\mid^{2}} &
            - c^{*}_{\mu}(x) \\
             c_{\mu}(x)          & 
\sqrt{1-\mid c_{\mu}(x)\mid^{2}}
        \end{array}
  \right)
\left(
        \begin{array}{cc}
   e^{i\theta_{\mu}(x)} &
            0           \\
            0             
  &  e^{-i\theta_{\mu}(x)}      
        \end{array}
  \right),
\end{eqnarray}
where the Abelian angle variable $\theta_{\mu}(x)$ and the 
non-Abelian variable $c_{\mu}(x)$ are defined in terms of 
$U_{\mu}(x)$ as $\tan\theta_{\mu}(x) = 
U_{\mu}^{3}(x)/U_{\mu}^{0}(x)$, $
 c_{\mu}(x)e^{i\theta_{\mu}(x)} 
= [-U_{\mu}^{2}(x) + i U_{\mu}^{1}(x)]$. 
To clarify the contribution of Abelian components to
the static force, we consider the Abelian projection of 
full non-Abelian link variables $U_{\mu}$ to the 
Abelian ones $u_{\mu}$. This is tantamount to  
replacing, in a Yang-Mills vacuum configuration put 
into MA gauge, of the $SU(2)$ link variables 
by $U(1)$ link variables, $U_{\mu}(x) \rightarrow u_{\mu}(x)$. 
Before this is done the off-diagonal parts $U_{\mu}^{1}(x)$ 
and $U_{\mu}^{2}(x)$ of gluon fields have been minimized 
by the gauge transformation which has to be found iteratively. 

One can further decompose the diagonal gluon component $\theta_{\mu}$ 
into the monopole part $\theta^{mo}_{\mu}$ and the photon part 
$\theta^{ph}_{\mu}$. Using the forward derivative 
$\partial_{\mu}f(x)\equiv f(x+\hat{\mu}a)-f(x)$, 
the 2-form $\theta_{\mu\nu}(x) \equiv \partial_{\mu}
\theta_{\nu}(x)-\partial_{\nu}\theta_{\mu}(x)$
defines the field strength which is separated as follows 
\begin{equation}
\theta_{\mu\nu}(x) = \bar{\theta}_{\mu\nu}(x) + 2\pi n_{\mu\nu}(x)
\label{eq:AFS}
\end{equation}
into a gauge invariant regular field strength 
$\bar{\theta}_{\mu\nu}(x) \equiv {\rm mod}_{2\pi}
\theta_{\mu\nu}(x) \in(-\pi, \pi]$ and a Dirac string 
part $n_{\mu\nu}(x) \in {\bf Z}$. 

From each part of the field strength the photon part 
$\theta^{ph}_{\mu}(x)$ and the monopole part 
$\theta^{mo}_{\mu}(x)$ of the $U(1)$ vector potential 
can be reconstructed, for instance 
\begin{equation}
\theta^{mo}_{\mu}(x) = 2\pi \sum_{x^{\prime}} 
\Box^{-1}(x-x^{\prime}) \partial_{\nu} 
n_{\mu\nu}(x^{\prime}) , 
\end{equation}
using the lattice Coulomb 
propagator $ \Box^{-1}=(\partial_{\mu}\partial^{\prime}_{\mu})^{-1}$,
where $\partial^{\prime}_{\mu}$ denotes the backward 
derivative. Then, we can construct the Abelian projected Wilson loop as  
\begin{equation}
\label{eq:AbelW}
\langle W_{abel}(C) \rangle = \langle \exp [ i 
\oint_{C}~\theta_{\mu}(x)~dx_{\mu} ] \rangle   
\end{equation}
which contains, besides of a ''photonic'' Wilson loop 
the monopole projected Wilson loop 
\begin{equation}
\label{eq:monoW}
\langle W_{mon}(C) \rangle = \langle \exp [ i 
\oint_{c}~\theta^{mo}_{\mu}~dx_{\mu} ] \rangle 
\end{equation}
as a uniquely defined factor.

On the other hand, we can also consider the center 
projection of an instanton configuration. This 
requires to go through the lattice discretization, too. 
Then, we use the direct maximal center gauge 
~\cite{DCG}. This gauge is defined as the 
gauge which brings directly the full link variables 
$U_l$ as close as possible to the center elements 
$\pm I$ by maximizing 
\begin{equation}
R_{MC}=\sum_{x,\mu}~[{\rm tr}~U_{\mu}(x)]^{2} .
\end{equation}
In this gauge the center projection is defined by 
\begin{equation}
Z_{\mu}(x) = {\rm sign} [{\rm tr}~U_{\mu}(x)] .
\end{equation}
After the gauge fixing, there remains only a local 
$Z(2)$ symmetry as $U_{\mu}(x) \rightarrow z(x) 
z(x+\hat{\mu}a) U_{\mu}(x)$ with $z(x) \in Z(2)$ 
Then, we can construct the center projected Wilson loop as  
\begin{equation}
\langle W_{center}(C) \rangle  = \langle \prod_{l \in C} Z_l \rangle .
\end{equation} 

\begin{figure}[bp]
\begin{center}
\begin{minipage}[hbt]{6cm}
\includegraphics[width=6cm]{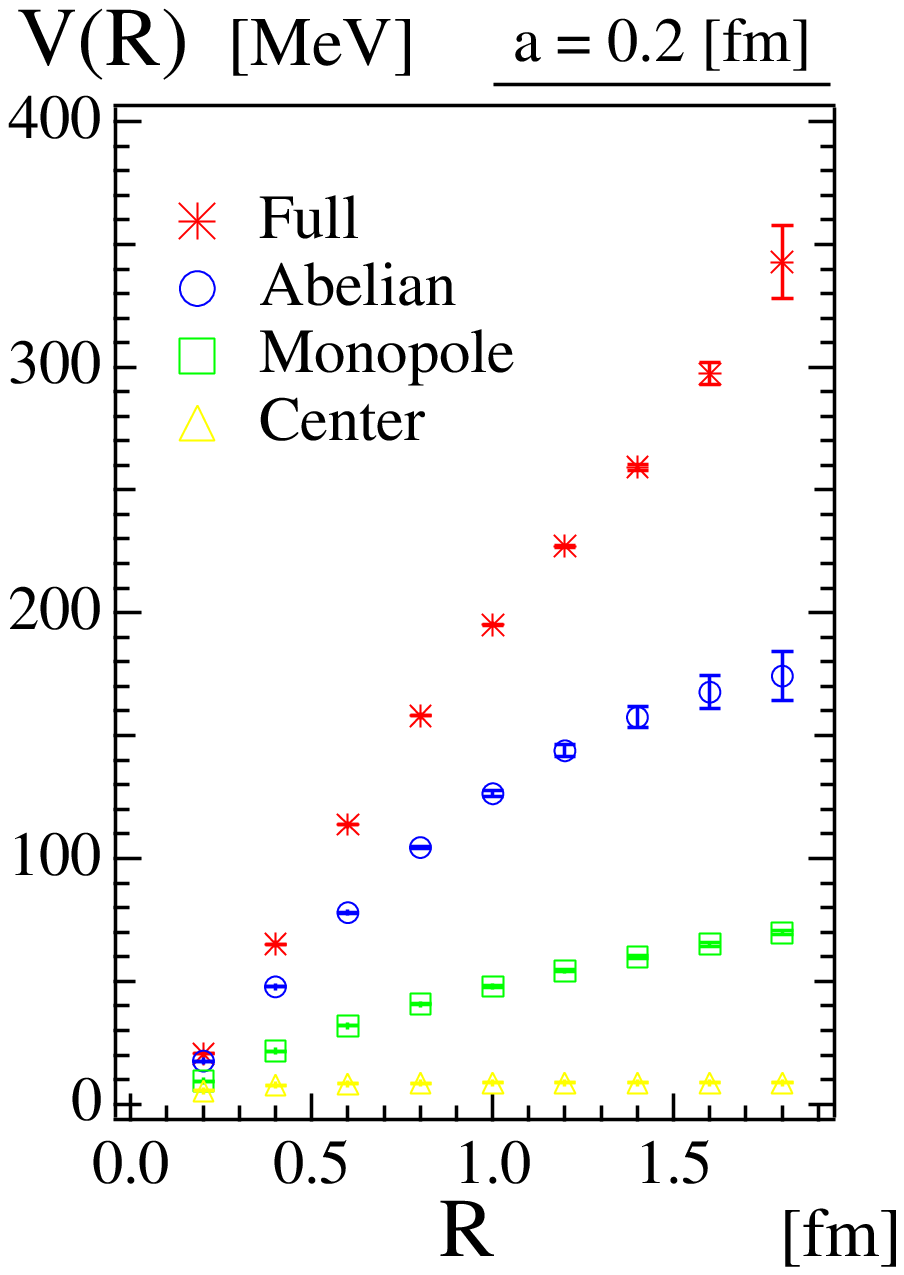}
\vspace*{-0.3cm}
\center{(a)}
\end{minipage}
\begin{minipage}[hbt]{6cm}
\includegraphics[width=6cm]{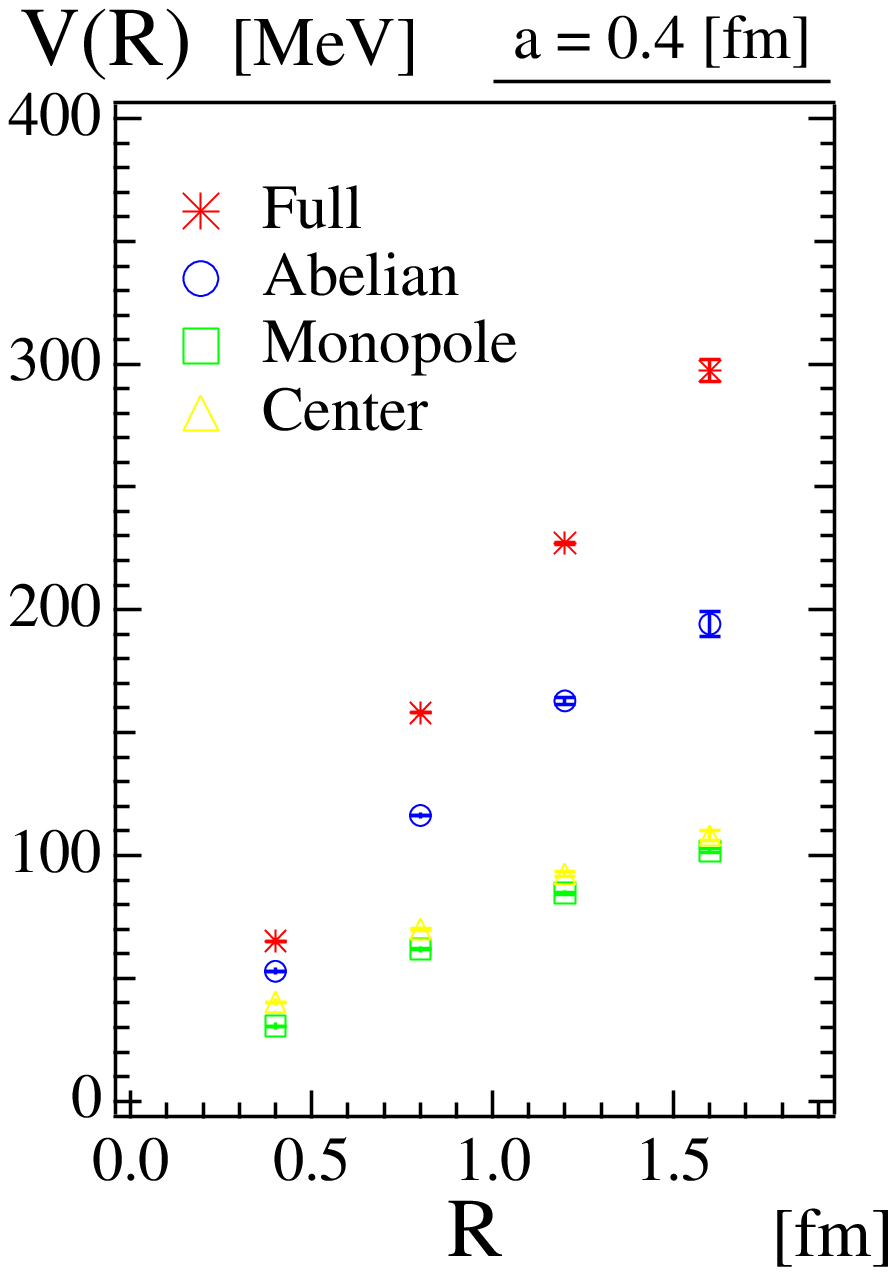}
\vspace*{-0.3cm}
\center{(b)}
\end{minipage}

\begin{minipage}[hbt]{6cm}
\includegraphics[width=6cm]{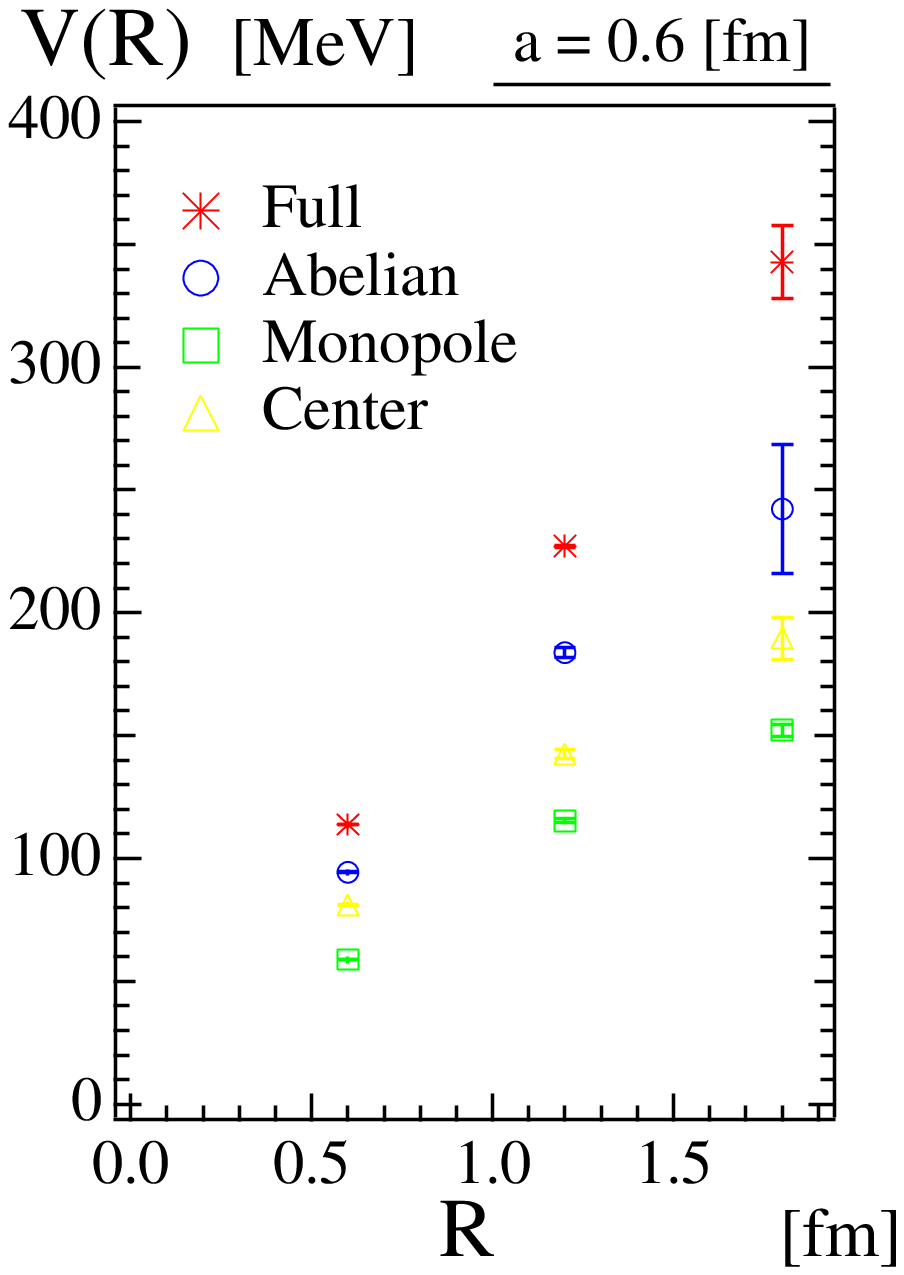}
\vspace*{-0.3cm}
\center{(c)}
\end{minipage}
\begin{minipage}[hbt]{6cm}
\includegraphics[width=6cm]{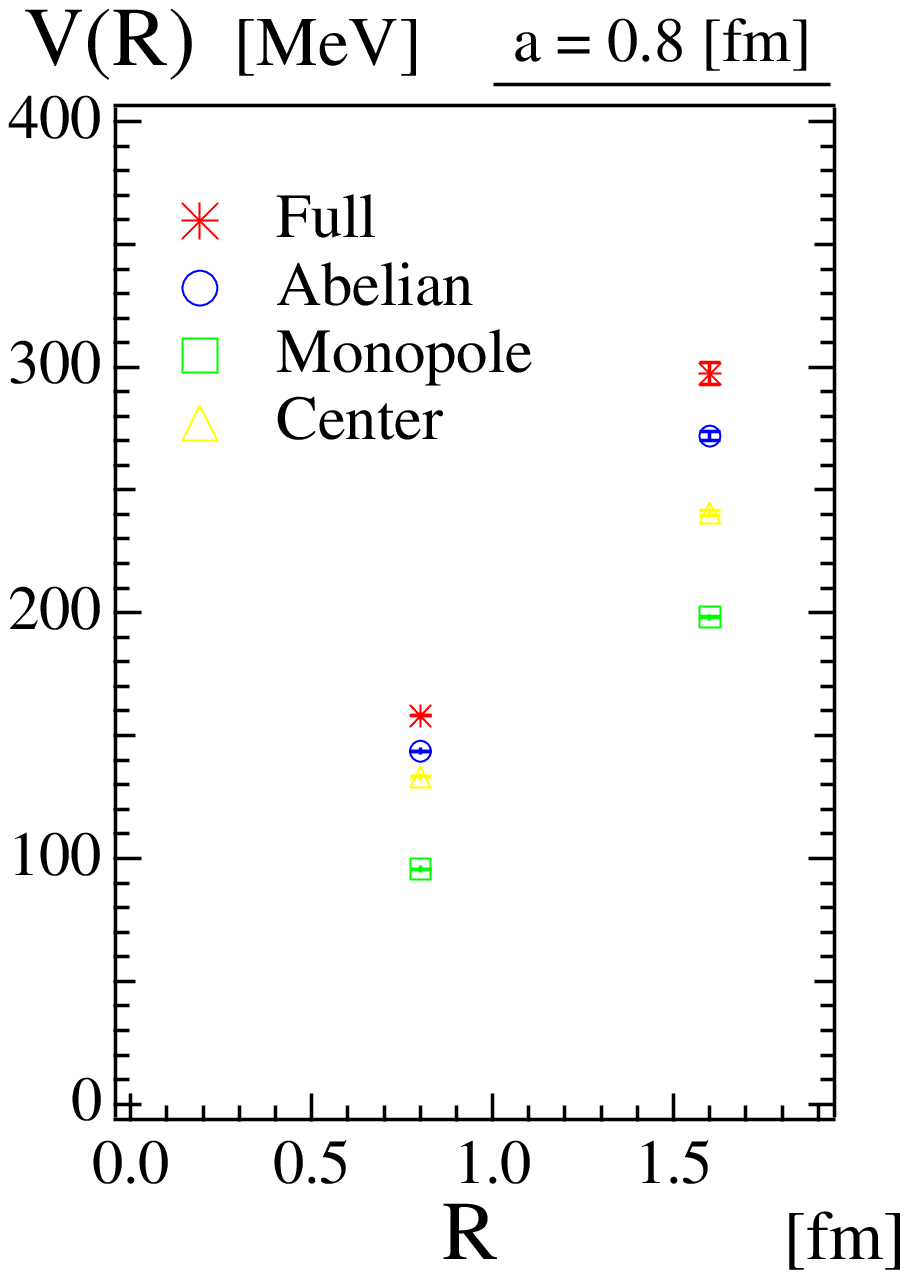}
\vspace*{-0.3cm}
\center{(d)}
\end{minipage}
\end{center}

\caption{
The non-Abelian static potential of fundamental charges 
compared with the Abelian projection, its monopole component 
and with the potential in $Z(2)$ center projection, for different
lattice spacings 
(a) $a=0.20~{\rm fm}$, 
(b) $a=0.40~{\rm fm}$, 
(c) $a=0.60~{\rm fm}$ and 
(d) $a=0.80~{\rm fm}$.  
The instanton size and density is fixed to 
$\bar{\rho}=0.4~{\rm fm}$ and $N/V=1~{\rm fm}$.}
\label{fig:FAPP}
\end{figure}

In Fig. \ref{fig:FAPP}, we show the instanton mediated 
non-perturbative force for the standard density 
$1~{\rm fm}^{-4}$ and fixed average instanton size 
$\bar{\rho}=0.4~{\rm fm}$. Instanton sizes were 
sampled according to the exponentially damped size 
distribution (\ref{eq:SizeDist}). The simulation data 
are based on a statistics of 1000 multiinstanton 
configurations for each discretization scale.

As shown in Fig. \ref{fig:FAPP}(a), for a lattice spacing 
much smaller than the typical size of instantons, the Abelian 
force calculated after Abelian projection from (\ref{eq:AbelW})
does not reproduce the non-Abelian force. Moreover, the Abelian 
force is practically reproduced after the monopole component 
of the Abelian field has been removed (by the ''photonic'' 
Wilson loops alone). The monopole contribution to the heavy 
charge force turns out to be smaller by a factor of two. 
The $Z(2)$ force calculated within the center-projected 
configurations is completely negligible.

For comparison we show in Fig. \ref{fig:FAPP}(b) the same 
for a coarser lattice of lattice spacing $a= 0.4~{\rm fm}
=\bar{\rho}$. In this case, the Abelian force has increased
but it is still far from reproducing the non-Abelian force.
The center component of the force has become measurable and
is comparable with the monopole component which has also 
increased somewhat.     

This trend continues when we consider a even coarser lattice
in Fig. \ref{fig:FAPP}(c) with $a= 0.6~{\rm fm} =1.5\bar{\rho}$.
In Fig.  \ref{fig:FAPP}(d) we show the coarsest lattice with 
$a= 0.8~{\rm fm} =2\bar{\rho}$. Now the slopes of the 
potentials calculated in the various projections are
in agreement with Abelian, monopole and center dominance,
with an ordering of the quasi-string tensions
$\sigma_{\mathrm{SU(2)}} > \sigma_{\mathrm{Abelian}} > 
\sigma_{\mathrm{Z(2)}} \approx \sigma_{\mathrm{mono}}$.
Here, of course, the potential can be looked at only at
one or two lattice spacings. 
Therefore the values of the string tension should be considered
only as rough estimates.

\section{Monopole Current Percolation}
\label{sec:sec5}

The monopole current can be defined using $u_{\mu}(x)$ 
following DeGrand and Toussaint~\cite{DeGrand+Toussaint}.
Since the Bianchi identity regarding the Abelian field 
strength $\bar{\theta}_{\mu\nu}(x)$ is broken by the 
decomposition in eq. (\ref{eq:AFS}), a monopole current 
$k_{\mu}(^*\!x)$ can be defined on the dual link $\{^*\!x,\mu\}$ as
\begin{equation}
k_{\mu}(^*\!x) \equiv \frac{1}{4\pi}\varepsilon_{\mu\nu\alpha\beta}
\partial_{\nu}\bar{\theta}_{\alpha\beta}(x+\hat{\mu}a)
=  - \partial_{\nu}\tilde{n}_{\mu\nu}(^*\!x)
\end{equation}
where $\tilde{n}_{\mu\nu}(x) \equiv \frac{1}{2}
\varepsilon_{\mu\nu\alpha\beta}n_{\alpha\beta}
(x+\hat{\mu}a)$. This current is obviously conserved, 
$\partial^{\prime}_{\mu}k_{\mu}(^*\!s) = 0$ which results 
in closed monopole loops on the 4-dimensional dual lattice. 
Here, $\partial^{\prime}_{\mu}$ denotes the backward derivative.

In the previous paper~\cite{Fukushima3}, ``block-spin'' 
transformations (of type II) have been applied to the monopole currents
~\cite{Ivanenko,Suzuki} in order to study how 
the global structure of the monopole loops changes for Abelian 
projected multi-instanton configurations. For large-scale 
blocking, leading to monopole currents defined on a very 
coarse (infrared) lattice, monopole clustering appears, 
i.e. clusters of big monopole length are seen to form.
  
\begin{figure}[h]
\begin{center}

\begin{minipage}[hbt]{6cm}
\includegraphics[width=6cm]{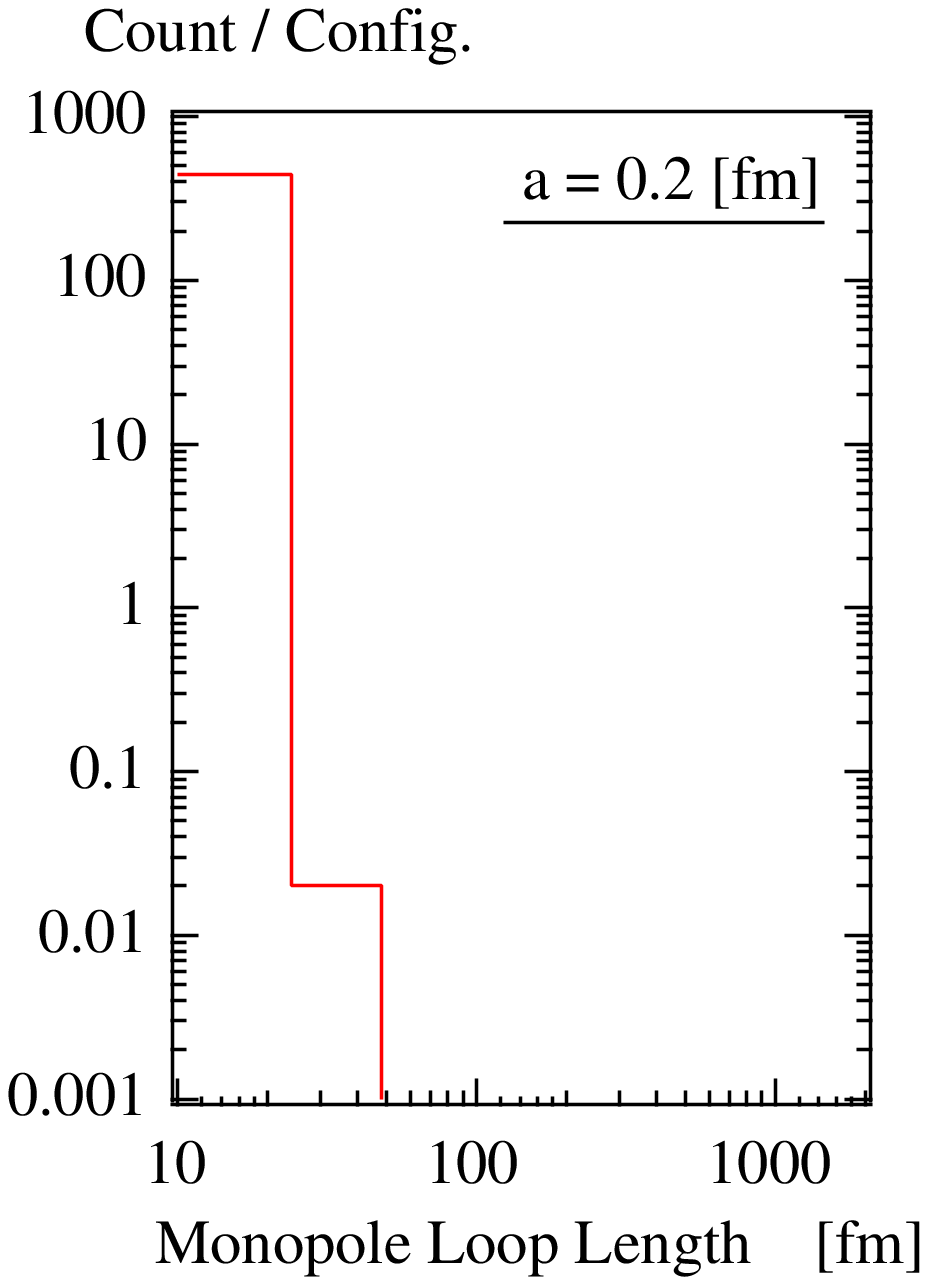}
\vspace*{-0.3cm}
\center{(a)}
\end{minipage}
\begin{minipage}[hbt]{6cm}
\includegraphics[width=6cm]{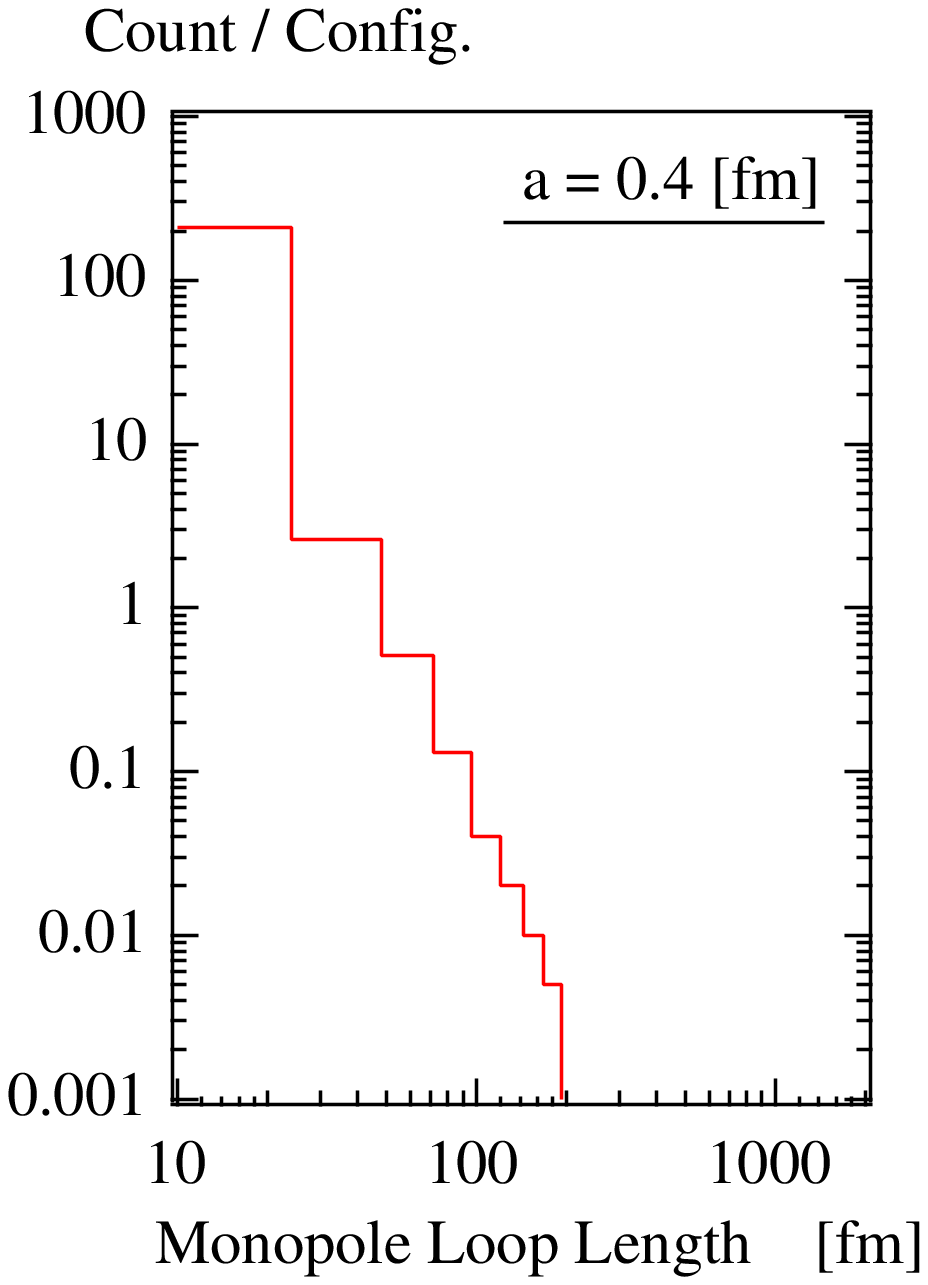}
\vspace*{-0.3cm}
\center{(b)}
\end{minipage}

\begin{minipage}[hbt]{6cm}
\includegraphics[width=6cm]{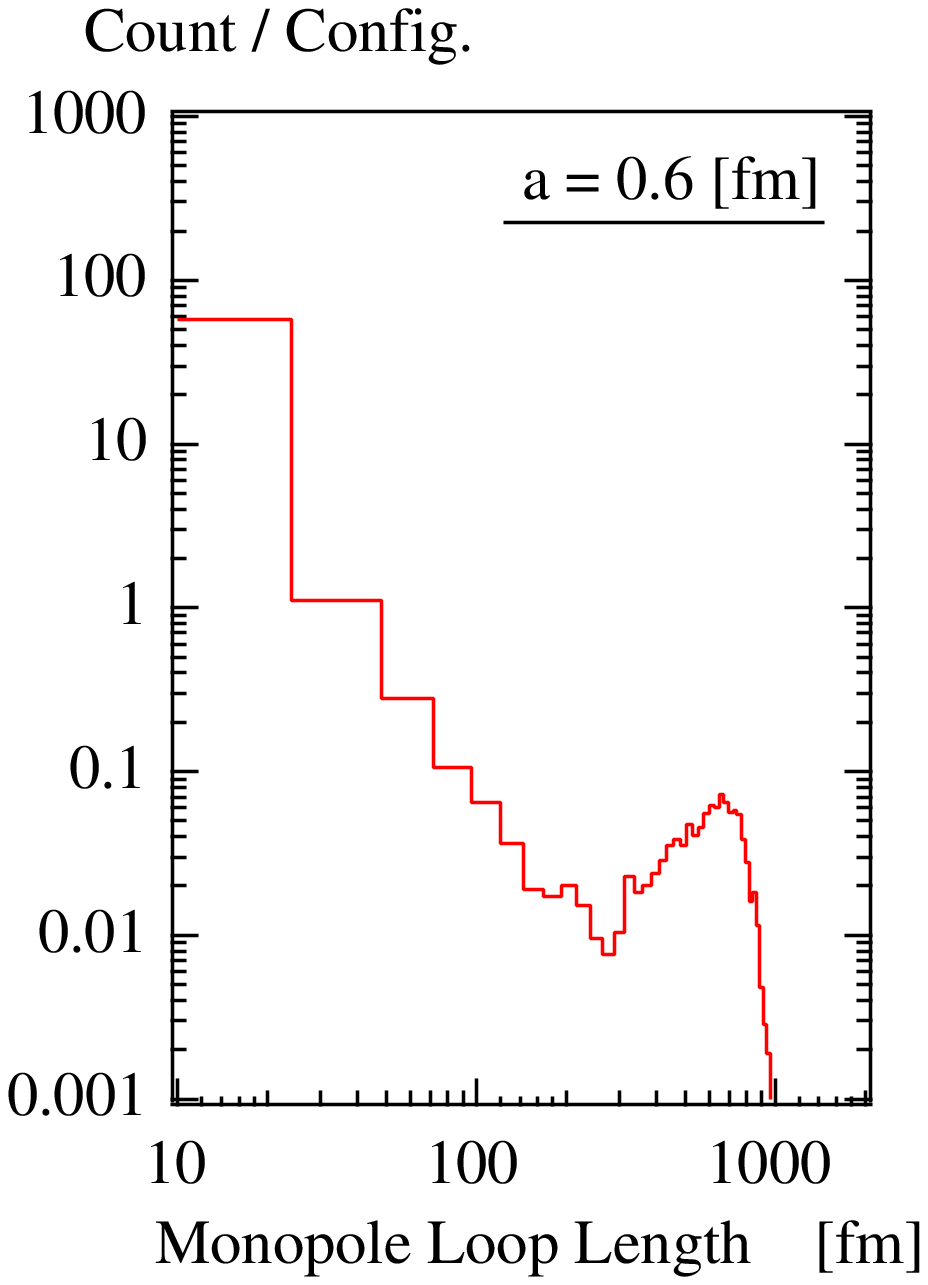}
\vspace*{-0.3cm}
\center{(c)}
\end{minipage}
\begin{minipage}[hbt]{6cm}
\includegraphics[width=6cm]{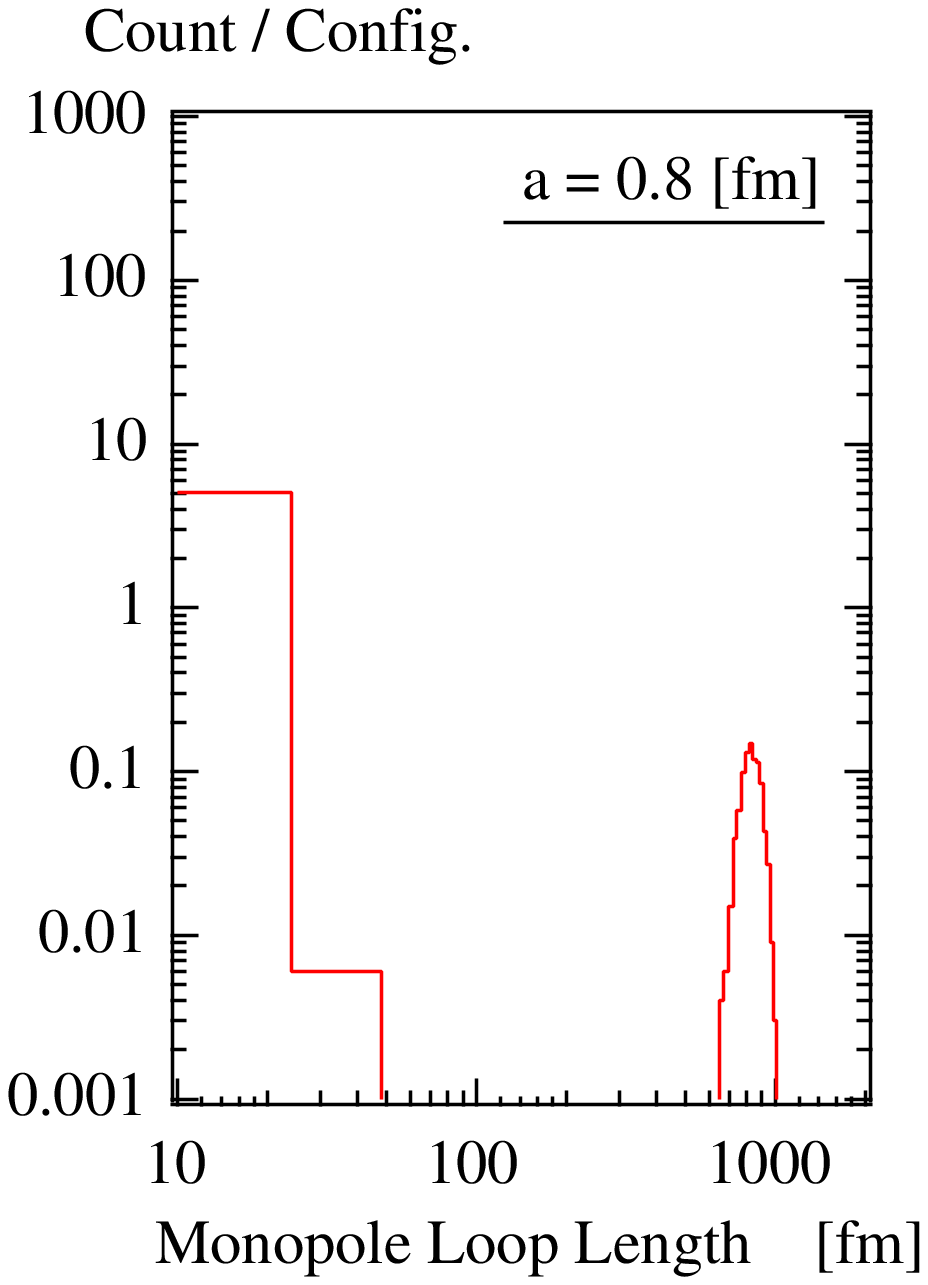}
\vspace*{-0.3cm}
\center{(d)}
\end{minipage}
\end{center}

\caption{
Histograms of connected monopole clusters per configuration
with respect to their lengths shown for different discretization
scales 
(a) $a=0.20~{\rm fm}$ 
(b) $a=0.40~{\rm fm}$, 
(c) $a=0.60~{\rm fm}$ and 
(d) $a=0.80~{\rm fm}$.}
\label{fig:HGRM}
\end{figure}

Instead of this, in the present study, we consider first 
the Abelian projection performed on lattices with different 
lattice spacings. The different lattices would be related, 
in the sense of \cite{Ivanenko} by type I
block spin transformations. For us, the difference is just in the
choice of the discretization scale for our multiinstanton configurations.
The histograms have been, however, obtained in independent runs.
In the case of Figs. \ref{fig:HGRM} (a) and (b), 2000 multiinstanton
configurations have been evaluated, in the case of 
Figs. \ref{fig:HGRM} (c) and (d) the statistics was 500 configurations each.
The Figs. \ref{fig:HGRM} show histograms $dN/dl$ 
monopole clusters normalized to the average occurrence in one configuration. 
The integral over lengths, $\int dN(l)$, gives the average number of 
separate clusters per configuration, while $\int l~dN(l)$ gives the 
average total length. When the projection is performed on a 
lattice with $a << \bar{\rho}$ (in our case $a=\frac12~\bar{\rho}$) 
the histogram of monopole loops per configuration with respect 
to their loop length (cluster size) contains only a component 
of relatively short loops as shown in Fig. \ref{fig:HGRM}(a). 
If the projection is done on a lattice with $a = 2\bar{\rho}$ 
the histogram Fig. \ref{fig:HGRM}(c) has 
already a component of monopole currents which percolate.
One very long monopole cluster of complicated structure
appears per configuration with a discretization scale $a=2\bar{\rho}$, 
similar to what has been observed previously ~\cite{Fukushima3}, 
as shown in Fig. \ref{fig:HGRM}(d).

\begin{figure}[h]
\begin{center}
\includegraphics[width=12cm]{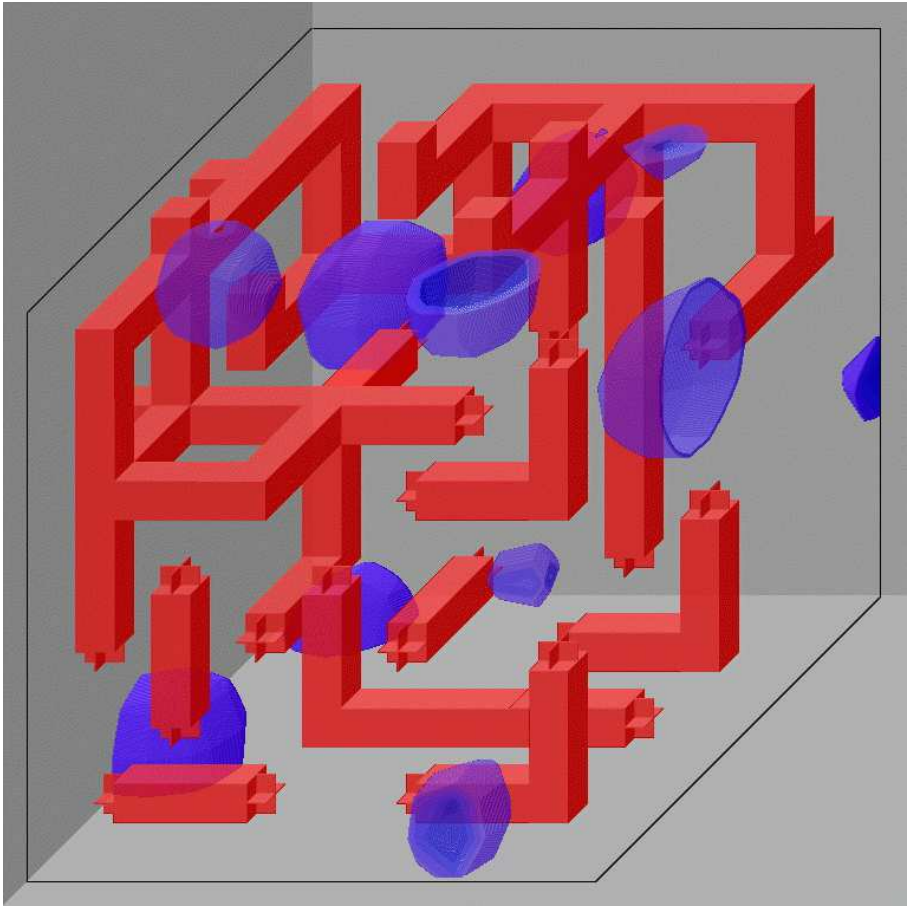}
\end{center}

\caption{
Glance into a time-slice of a typical 
configuration 
with continuum instantons (balls)
and lattice monopoles (bars) living on the 
coarsest lattice (with lattice spacing 
$a=0.8~{\rm fm}$). 
Only a subvolume $V=(3.2~{\rm fm})^3$ 
is shown. } 
\label{fig:VISU}
\end{figure}

In Fig. \ref{fig:VISU} we show, in a 3-dimensional cut,
the instanton and monopole content of a characteristic 
configuration at that density. The cross-shaped endcaps 
of the bars symbolize that the monopole world lines 
leave (enter) the volume going to (coming from) another 
timeslice. In MAG, the monopole network (percolating at
the coarsest lattice) is wrapping the instantons.

\section{Summary and Conclusions}
\label{sec:sec6}

We have studied the static quark potential induced by a random 
instanton liquid. We have made the present study in the sequel 
of the previous work~\cite{Fukushima3} where an infrared 
suppression of the instanton size distribution $dn(\rho) 
\propto \rho^{-\nu}~d\rho$ has been assumed.
In this paper, we adopt a size distribution as $dn(\rho)/d\rho 
= \alpha\rho^{b-5} \exp(-\beta\rho^2/ \overline{\rho^2})$ 
(suggested by lattice instanton searches) adapted to a space-time 
density of $N/V =1~{\rm fm}^{-4}$ and an average instanton size 
$\bar{\rho} =0.4~{\rm fm}$. We find that the gross effects do not 
depend on the detailed form of the size distribution. 

First, we have compared the non-Abelian potential inferred from 
the Wilson loop embedded in the simulated random multi-instanton 
ensemble with the well-known dilute-gas expression which reduces 
the potential to the effect of isolated single instantons.
We found that the dilute gas expression differs essentially from
the simulation results if the density is bigger than $\frac{1}{20}$
of the phenomenological value, due to many instantons interacting 
with a Wilson loop in a non-exponentiable way.  This explains why
the quasi-linear part of the $\bar{Q}Q$ force at 
$R \sim 1 - 1.5~{\rm fm}$
needs to be found from a simulation. We have found that 
the force between adjoint charges behaves more complicated in
the multi-instanton system. 

Second, we have investigated the Abelian and center projection 
of the gauge field configurations built up by random instanton 
configuration. The gauge fixing and subsequent projection has 
been performed introducing
lattices of different lattice spacing, and we have 
studied the effect of changing the discretization scale, in 
particular checking lattice spacings $a=0.2$, $0.4$, $0.6$ and $0.8~{\rm fm}$. 
In the finer lattice case, the monopole contribution to the 
heavy charge force turns out to be much smaller than the Abelian 
projected force. The force calculated for the center-projected 
configurations is completely negligible. With larger and larger 
discretization scale, Abelian, monopole and center dominance 
becomes restored, which is almost perfectly illustrated for 
$a=2\bar{\rho}$. Then the quasi-linear non-Abelian force 
can be almost reproduced by the corresponding projected Wilson loops.

We stress this result, although the random instanton 
liquid model as such does not appear to be a realistic model
for the confining aspect of the Yang-Mills vacuum, because this seems
to reflect a more general feature to be kept in mind for more 
sophisticated semiclassical models. 

We see that in the multi-instanton liquid, at $a=2~\bar{\rho}$, 
Abelian dominance of the heavy charge force amounts to about 90 \%, 
which is mainly due to the Abelian monopole component.
The singular part of the Abelian gauge potential accounts
for 80 \% of the Abelian force. The center projected force now
also describes about 80 \% of the non-Abelian one.

We have considered the monopole loop percolation behavior in the 
random multi-instanton configurations using the same typical lattice 
spacings. Only after performing the projection with a 
discretization scale $a>\bar{\rho}$ we could find a percolating 
component among the monopole currents, represented by one or few clusters of 
huge loop length per configuration. This parallels the previous 
observation in Ref. \cite{Fukushima3} that blocking of monopole currents 
is necessary to see percolation/condensation happening. 

\section*{Acknowledgment}
E.-M.~I. gratefully acknowledges the support by the
Ministry of Education, Culture and Science of Japan (Monbu-sho)
providing the opportunity to work at RCNP. M.~F. is supported by JAERI as 
a research fellow.

\clearpage

\section{Figure Caption}

\clearpage

\clearpage

\clearpage

\clearpage

\clearpage

\clearpage

\end{document}